\documentclass[preprint,pre,superscriptaddress,showpacs,showkeys]{revtex4}

\usepackage{graphicx}
\usepackage{amsmath}
\usepackage{amssymb}
\usepackage{amsfonts}

\begin{document}

\title[2D flow of foam: local measurements]{Two-dimensional flow of foam around a circular obstacle: local measurements
of elasticity, plasticity and flow}

\author{Benjamin Dollet}
\email{b.dollet@utwente.nl} \altaffiliation{Present address:
Physics of Fluids, University of Twente, PO Box 217, 7500 AE
Enschede, The Netherlands.}
\author{Fran\c cois Graner}
\affiliation{Laboratoire de Spectrom\'etrie Physique, BP 87, 38402
Saint-Martin-d'H\`eres Cedex, France}

%

\begin{abstract}
We investigate the two-dimensional flow of a liquid foam around
circular obstacles by measuring all the local fields necessary to
describe this flow: velocity, pressure, bubble deformations and
rearrangements. We show how our experimental setup, a quasi-2D
"liquid pool" system, is adapted to the determination of these
fields: the velocity and bubble deformations are easy to measure
from 2D movies, and the pressure can be measured by exploiting a
specific feature of this system, a 2D effective compressibility.
To describe accurately bubble rearrangements, we propose a new,
tensorial descriptor. All these quantities are evaluated via an
averaging procedure that we justify showing that the fluctuations
of the fields are essentially random. The flow is extensively
studied in a reference experimental case; the velocity presents an
overshoot in the wake of the obstacle, the pressure is maximum at
the leading side and minimal at the trailing side. The study of
the elastic deformations and of the velocity gradients shows that
the transition between plug flow and yielded regions is smooth.
Our tensorial description of T1s highlight their correlation both
with the bubble deformations and the velocity gradients. A salient
feature of the flow, notably on the velocity and T1 repartition,
is a marked asymmetry upstream/downstream, signature of the
elastic behaviour of the foam. We show that the results do not
change qualitatively when various control parameters (flow rate,
bubble area, fluid fraction, bulk viscosity, obstacle size and
boundary conditions) vary, identifying a robust quasistatic
regime. These results are discussed in the frame of the actual
foam rheology literature, and we argue that they constitute a
severe test for existing rheological models, since they capture
both the elastic, plastic and fluid behaviour of the foam.
\end{abstract}

\maketitle

%

%

\section{Introduction}

Liquid foams have a ubiquitous mechanical behavior: depending on
the strength of an external applied solicitation, they can exhibit
both elastic, plastic or viscous response
(\cite{Weaire1999,Hohler2005}). This complex behavior is used in
many industrial applications (\cite{Khan1996}), like ore
flotation, oil extraction, food and cosmetic industry. Liquid
foams are also of fundamental interest as models to study complex
fluids, since their constitutive item, the bubble, is
experimentally easily observable, contrary to colloids or
polymers. The understanding of foam rheology has motivated active
research; a series of seminal studies has first focused on the
elastic properties of foams, such as elastic moduli or yield
stress (\cite{Derjaguin1933,Princen1983,Stamenovic1984,Khan1986}).
The plasticity of foams originates from topological
rearrangements, called T1s (Fig. \ref{Skeletons}c). This coupling
between local rearrangements of constitutive entities and a
macroscopic plastic behaviour is a general feature of many
materials (emulsions, pastes, slurries), that studies have
described generically as soft glassy materials
(\cite{Sollich1997}). On the other hand, the research on the
viscous and dissipative properties of foams is rather recent
(\cite{Kern2004,Denkov2005,Cantat2005,Dollet2005c}).

Much effort is currently devoted to integrate elastic, plastic and
viscous behaviours in a single constitutive equation
(\cite{Hohler2005,Weaire,Graner}). To achieve such a goal, a fine
and precise knowledge of the mechanical behavior of the foam is
required. This is the reason why foams are often studied in
quasi-2D geometries (\cite{Cox2003,Vaz2005}), where foams are
confined so that they are only one bubble thick (confinement
either between two parallel plates, between one horizontal plate
and the surface of a soap solution, or at the free surface of a
soap solution). Imaging is easy in these quasi-2D geometries,
contrary to an opaque 3D foam. A classical way to study quasi-2D
foams consists in using rheometric flows (\cite{Larson1999}), like
simple shear flows (\cite{Debregeas2001,Lauridsen2002}), which are
easy to analyze. The study of heterogeneous flows is
complementary: it is less easy to analyze and to understand, but
it enlarges the number of observed effects. This is the case for
example for flows in constrictions (\cite{Asipauskas2003}), or for
Stokes experiments, i.e. flows around obstacles
(\cite{Alonso2000,deBruyn2004,Dollet2005a,Cantat2005b,Dollet2005b,DolletEllipse}).

We studied extensively Stokes experiments for foams, focusing on
the effect of the foam on the obstacle: drag (\cite{Dollet2005a}),
lift (\cite{Dollet2005b}) or torque (\cite{DolletEllipse}). The
variations of these quantities with various control parameters,
especially the foam flow rate, illustrate the interplay between
elastic, plastic and viscous behavior of the foam. To go beyond
these force and torque measurements, we revisit here some of these
experiments with the complementary approach, studying by local
measurements the effect of the presence of an obstacle on a
flowing foam. We present a method developed to analyze precisely
and completely this local response, quantified by various local
fields. We extract velocity and velocity gradients, bubble
deformations, and pressure, which describe the elastic and viscous
parts of the foam response. We also define a new tensorial
descriptor of the bubble rearrangements, and present the
corresponding field, which quantifies accurately foam plastic
response. The results highlight the differences of the foam local
response with respect to simple viscoplastic and viscoelastic
responses, namely a marked up/downstream asymmetry, signature of a
delayed, elastic response of the foam, and a strong plastic
response wherever the bubble deformation becomes close to a
critical, yield strain. Our results thus call for a modelisation
coupling elastic, plastic and fluid behavior, and they also
constitute an extensive database to test and constrain such
models.

%

%
%
%

%
%

\section{Materials and methods}

\subsection{Experiment}

\begin{figure}
\begin{center}
\includegraphics[width=12cm]{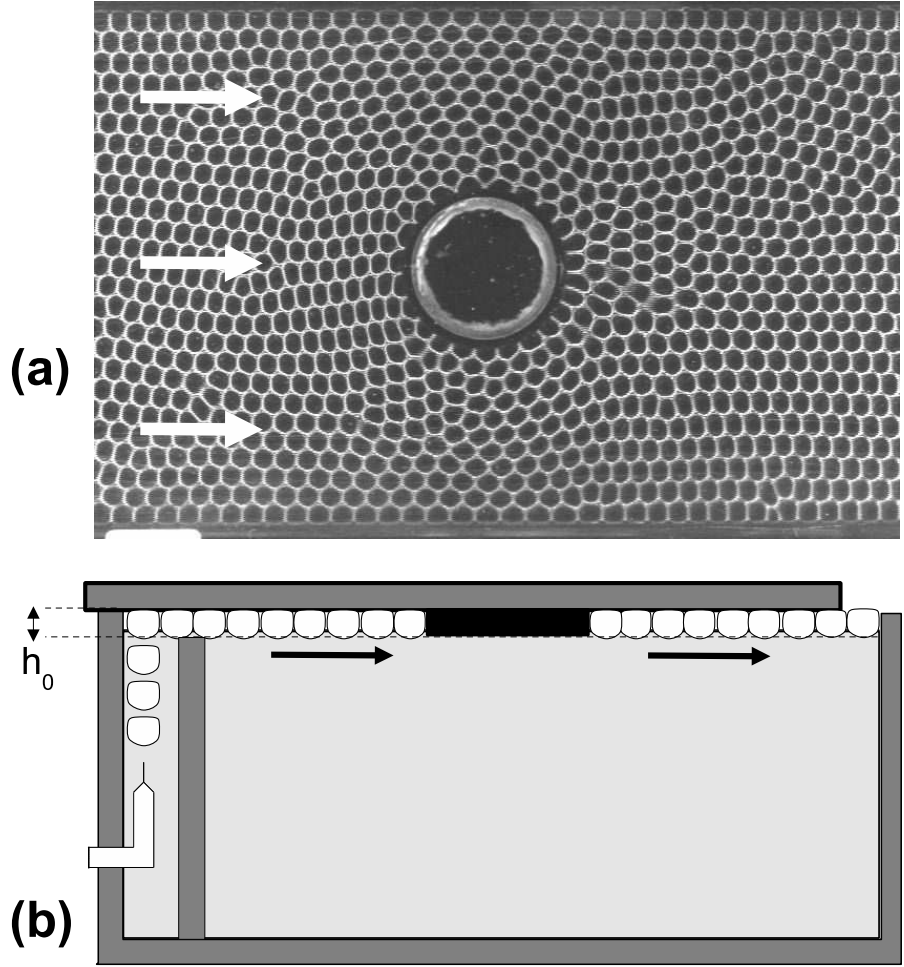}
\caption{\label{Image} (a) Photo of foam flowing from left to
right (arrows) around a circular obstacle of diameter 30 mm. The
bubble size is 16.0 mm$^2$ (note the monodispersity of the foam),
and the flow rate is 176 ml min$^{-1}$. The walls of the channel
(width 10 cm) are visible at the top and bottom of the picture.
The surface of the observed field is $15.4\times 10.2$ cm$^2$.
Movies are available at
\texttt{http://www-lsp.ujf-grenoble.fr/recherche/a3t2/a3t2a1/mousses2d3d.htm}.
(b) Side view of the setup. The foam is constituted by a monolayer
of bubbles and the black rectangle represents the
obstacle.}\end{center}
\end{figure}

We perform a Stokes experiment
(\cite{deBruyn2004,Dollet2005a,Alonso2000,Asipauskas2003}), i.e.
we study the flow of foam around obstacles (Fig. \ref{Image}),
using a foam channel fully described in \cite{Dollet2005a}.
Briefly, a tank is filled with a bulk solution obtained by adding
1\% of commercial dish-washing liquid (Taci, Henkel) to desionised
water. Its surface tension, measured with the oscillating bubble
method, is $\gamma = 26.1\pm 0.2$ mN m$^{-1}$, and its kinematic
viscosity, measured with a capillary viscosimeter, is $1.06\pm
0.04$ mm s$^{-2}$ unless explicitly stated. Nitrogen is blown in
the solution through a nozzle or a tube at a computer controlled
flow rate. This generates a "liquid pool foam" foam
\cite{Vaz2005}, constituted by a horizontal monolayer of
monodisperse bubbles (dispersity $<3\%$) of average thickness
$h_0$, confined between the bulk solution and a glass top plate.
This is a quasi-2D foam (\cite{Cox2003,Vaz2005}): despite the 3D
geometry of the bubbles (Fig. \ref{Image}b), it experiences 2D
horizontal flows, which minimizes the effect of drainage. Two
others quasi-2D foams exist: the bubble raft (no confinement), and
the Hele-Shaw cell (confinement between two horizontal plates).
Contrary to these two systems, the liquid pool foam has an
effective in-plane compressibility (\cite{Dollet2005a}), which
enables to measure pressure easily, as recalled in Section
\ref{SectionPressure}. The fluid fraction is adjusted by the foam
thickness \cite{Raufaste}; its value is 7\%. The foam flows around
an obstacle placed at the middle of the channel; in the present
study, we choose the flow rate between 24 and 515 ml min$^{-1}$
(corresponding velocities: 0.11 to 2.5 cm s$^{-1}$). In this
paper, we will study a reference case characterized by the
following values of the parameters: circular obstacle of diameter
30 mm, flow rate of 176 ml min$^{-1}$, bubble area of 16.0 mm$^2$,
foam thickness of 3.5 mm, and bulk viscosity of 1.06 mm$^2$
s$^{-1}$. We will then study the influence of each control
parameter separately.

This setup has allowed us to measure forces on obstacles
(\cite{Dollet2005a,Dollet2005b}) and pressure drops associated to
the flow of foam (\cite{Dollet2005c}). For the purpose of the
present paper, we record for every experiment a stack of 750
images, representing a movie of 30 seconds. From these movies, we
extract all relevant quantities describing the flow of foams:
velocity, pressure, elastic stress, bubble deformations, and
bubble swapping (topological rearrangements, or T1s) using a
home-made procedure, as follows.

\subsection{Image analysis}

\subsubsection{Skeletonisation of experimental images}

\begin{figure}
\begin{center}
\includegraphics[width=12cm]{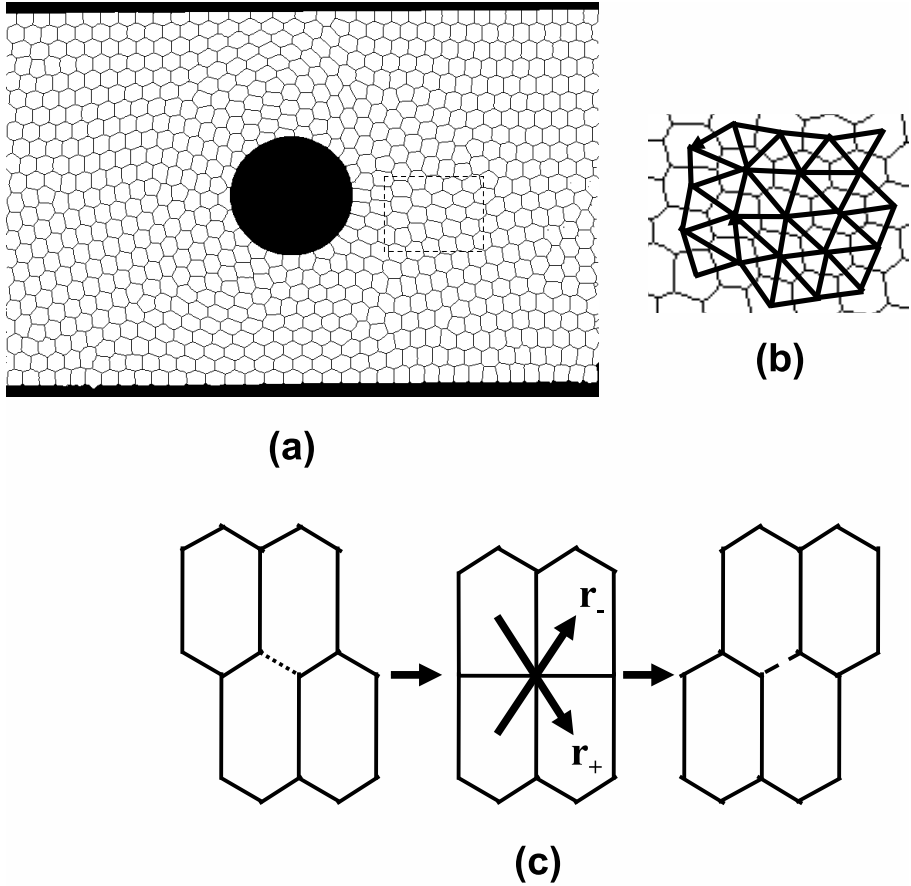}
\caption{\label{Skeletons} (a) Skeletonized image of foam. (b)
Zoom on the zone framed in (a): the network of bubble edges is
figured with thin lines, and the (triangular) center network with
thick lines. (c) Sketch of a side swapping (topological
rearrangements, also called T1 event). Left: the edge to disappear
is dotted; middle: definition of the vectors $\vec{r}_+$ and
$\vec{r}_-$; right: the new edge is dashed.}\end{center}
\end{figure}

With the NIH Image software, we invert the grey levels of the
images, then threshold our images, to clearly separate the black
network of edges from white bubbles. We have defined several zones
on the image each with different thresholds, to compensate from
slight remaining spatial variations of light intensity. We finally
extract the network of the bubble edges from the experimental
images by a classical skeletonisation procedure, which reduces the
foam to a network of one-pixel thick edges (Fig.
\ref{Skeletons}a).

This procedure conserves the topology between the real and the
skeletonized bubbles (Figs. \ref{Image}a and \ref{Skeletons}a),
which enables a proper evaluation of the bubble deformation, as
explained in Section \ref{TensorialFields}. It has two
limitations: first, it distorts the geometry and curvature of the
bubbles edges and vertices, which prevents us from evaluating
precisely the elastic stress, since this requires integration
along all edges (\cite{Batchelor1970}). Second, it is not adapted
to the boundaries; we therefore systematically eliminate the data
near the obstacle and the channel walls.

\subsubsection{Treatment of skeletonized images} \label{Treatment}

For a 2D skeletonized foam, the bubbles are bounded by thin edges,
which merge in threefold vertices. Bubbles are thus easily
labelized, and a vertex can be unambiguously defined as a black
pixel for which, among its eight neighboring pixels, one can find
three pixels belonging to three different bubbles. Boundaries
vertices are defined as pixels on boundaries, with two neighboring
pixels belonging to two different bubbles.

We scan an image in two steps. In a first step, each individual
bubble is labelized with a different number. The program records a
list of bubbles, each bubble being represented by its number $b$,
its number of pixels $N_b$, and the position $\vec{x}_b$ of its
barycenter. In a second step, vertices are identified and
labelized, and the program records them in a second list, each
vertex being represented by its number, its coordinates and the
label of its neighboring bubbles. The subsequent analysis does not
require the image.

\begin{figure}
\begin{center}
\includegraphics[width=12cm]{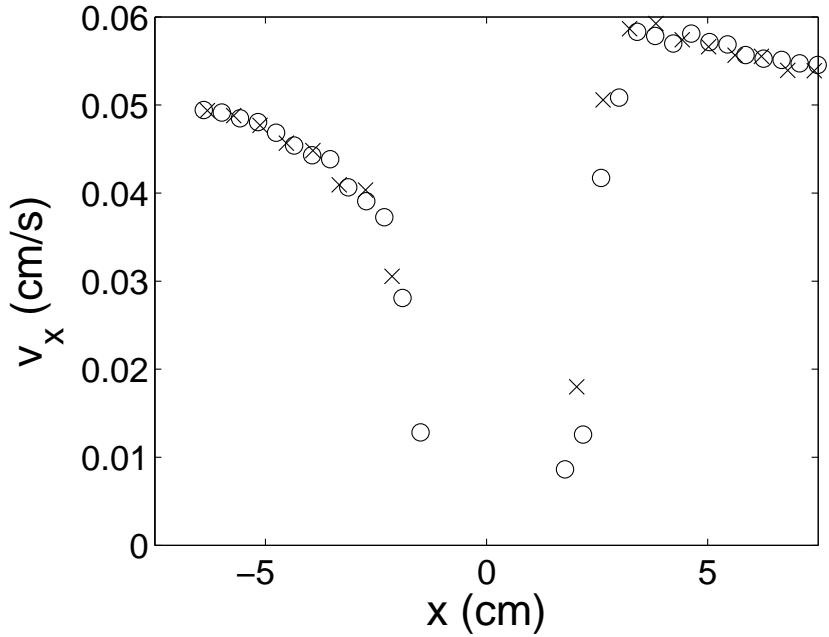}
\caption{\label{BoxSize} Plot of the velocity $v_x$ along the axis
of symmetry of the flow, as a function of the streamwise
coordinate $x$, for a flow rate of 24 ml min$^{-1}$ and for two
mesh sizes ($\times$: boxes of area $6.0\times 6.2$ mm$^2$,
$\circ$: boxes of area $4.1\times 4.1$ mm$^2$).}\end{center}
\end{figure}

To compute the fields, we mesh the image by a rectangular grid. We
have checked that there exists a mesh size for which the results
do not change (Fig. \ref{BoxSize}): this validates our choice, and
is a first indication of the continuous character of the foam (see
section \ref{AveragesFluctuations}). We have chosen to mesh the
image with a rectangular grid of $26\times 17$ (nearly) square
boxes of side 6 mm: such a choice enables us to capture well the
variations at the macroscopic scale, and the statistics is
sufficient for these variations to be smooth (during the whole
movie, about $2\times 10^3$ bubbles are computed per box). For
simplicity, each bubble is attributed to the box where its center
lies even if part of it belongs to other boxes.

\subsubsection{Direct measurements from skeletonized images}

From the list of bubbles, we compute the network of the vectors
$\vec{r}$ linking two centers of bubbles in contact, which we call
the center network (Fig. \ref{Skeletons}b). To be precise, each
vector $\vec{r}$ is attributed to the two boxes (with a
coefficient $1/2$ each) of the two bubble centers it binds. We
then compute the texture tensor (\cite{Aubouy2003}):
\begin{equation}\label{DefinitionM}
\bar{\bar{M}} = \langle \vec{r}\otimes\vec{r} \rangle ,
\end{equation}
    which
    is the second-order tensor of components: $M_{ij} = \langle r_i r_j \rangle$. The average is performed over
    750 images and all vectors in the box. This
    tensorial quantity has proven to be a good descriptor of
    bubble deformation: it reproduces the size, direction and amplitude of deformation of bubbles, in
    studies where it has been computed over the bubble edges network $\vec{\ell}$
    (\cite{Asipauskas2003,Courty2003,Janiaud2005}).
    Here, we calculate it over the center network. With such a
    choice,
    the texture tensor definition (\ref{DefinitionM}) is much more
    general than with the bubble edges network: it applies to 3D
    foams, and to wet foams such as the ones considered hereby. It
    is also more robust, because the center of masses, hence the
    vectors $\vec{r}$, is much less biased by skeletonisation than the bubble edges. Anyway,
     the two possible network
    choices are equivalent at low bubble deformation.

In order to compute the velocity field, we compare successive
frames. In the studied range of flow rates, the displacement of a
bubble between two successive frames is small compared to its
size; the displacement of the bubble centers is thus easy to
calculate, and we average all displacements on each box to get the
velocity field (Eulerian rather than Lagrangian point of view).

The calculation of the T1s is also based on the correlation of two
successive images; a T1 is a topological neighbor-swapping event,
during which a bubble edge disappears and a new one is created
(Fig. \ref{Skeletons}c). The program tracks independently the
disappearing and appearing edges, by comparing the list of edges
of two successive frames. This decoupling of the disappearing and
appearing edges is necessary for two reasons: first, the duration
of a T1 event may be longer than the time interval between two
successive frames (0.04 s); second, the transient fourfold vertex
(middle of Fig. \ref{Skeletons}c) contains a certain amount of
liquid. After skeletonisation, this is often erroneously
recognized as an artificial little four-sided bubble between the
four bubbles experiencing the T1, which we have to remove by
imposing a lower threshold on the bubble area. Actually, a T1
covers two distinct instantaneous events: one disappearance and
one apparition of a link between two bubbles. To a disappearing
(appearing) edge is associated the vector of the center network
$\vec{r}_-$ ($\vec{r}_+$) which links the centers of the two
separating (attaching) bubbles. A complete quantification of T1s,
including not only their frequency but also their direction, must
rely on these vectors, whose direction is irrelevant by
definition. We thus define the tensors $\bar{\bar{T}}_{\pm} =
f_{\pm} \langle \vec{r}_{\pm} \otimes \vec{r}_{\pm} \rangle$,
where $f_+$ ($f_-$) is the frequency of separation (attachment)
events per link $\vec{r}$ of the center network (\cite{Graner}).
Preliminary studies show that these tensors are closely related to
the mechanical properties of the foam (\cite{Graner}). However, in
this paper, we use a more intuitive definition, based on unit
vectors: $\bar{\bar{T}}^{\mathrm{adim}}_{\pm} = f_{\pm} \langle
\hat{r}_{\pm} \otimes \hat{r}_{\pm} \rangle$, because it has the
advantage to be directly proportional to the frequency of T1s.

\subsection{Computation of the fields}

We present here the relevant fields describing the flow of foams,
and the way they are computed from the image analysis detailed in
the previous section.

\subsubsection{Pressure} \label{SectionPressure}

As already mentioned in a previous study (\cite{Dollet2005a}), in
a quasi-2D setup with foam confined between a top plate and a
liquid pool, the depth of bubbles adjust to pressure variations.
The 3D compressibility of the bubbles is negligible here: for an
ideal, isothermal gas, the compression modulus is of order $10^5$
Pa, which is 4 orders of magnitude higher to the measured local
variations of pressure, as shown later (Fig. \ref{Pressure}).
Hence, the volume of a given bubble is constant; but if its
pressure increases, its depth increases to maintain equilibrium
with the hydrostatic pressure of the bulk solution, hence its
visible area decreases: the foam has an effective 2D
compressibility in the plane of the top plate. The precise
relationship between bubble area and pressure is established in
another paper (\cite{Raufaste}), in which we show that the
relative uncertainty in pressure equals 2\%. It writes:
\begin{equation}\label{P(A)_formula}
P - P_0(x) = \frac{\rho g\mathcal{V}}{A} +
2\gamma\sqrt{\frac{\pi}{A}} ,
\end{equation}
with $\rho = 10^3$ kg m$^{-3}$ the volumetric mass of the
solution, $g = 9.8$ m s$^{-2}$ the gravity acceleration, and
$\mathcal{V}$ the constant bubble volume. Here, $P_0$ is the local
reference pressure, which embodies the constant pressure gradient
along the channel (\cite{Dollet2005c}): hence, $P-P_0$ is the
local variation of pressure due to the presence of the obstacle.
The average bubble area is easily computed in each box with the
image analysis program. Since the pressure field is scalar, it is
convenient to represent it in gray levels.

\subsubsection{Velocity and velocity gradients}

The image analysis program provides directly the velocity field,
which we represent as usual with arrows. The velocity gradient is
computed by finite differences; we evaluate this gradient in the
middle of the four boxes $(i,j)$, $(i,j+1)$, $(i+1,j)$ and
$(i+1,j+1)$. 
We rather use the symmetric velocity gradient, the deformation
rate: $\bar{\bar{D}} = (\overline{\overline{\nabla v}} +
{^t}\overline{\overline{\nabla v}})/2$, and the antisymmetric
velocity gradient, the vorticity, which is a scalar for 2D flows:
$\omega = \dfrac{1}{2} \left( \dfrac{\partial v_y}{\partial x} -
\dfrac{\partial v_x}{\partial y} \right)$. We will precise the
graphical representation of the deformation rate, and of the
others tensorial quantities, in the following sections. We will
also use the scalar dissipation function (\cite{Guyon2001}),
defined as: $\| \bar{\bar{D}} \| = \sqrt{D_{xx}^2 + 2D_{xy}^2 +
D_{yy}^2}$.

\subsubsection{Tensorial fields: texture, statistical elastic strain,
T1} \label{TensorialFields}

\begin{figure}
\begin{center}
\includegraphics[width=12cm]{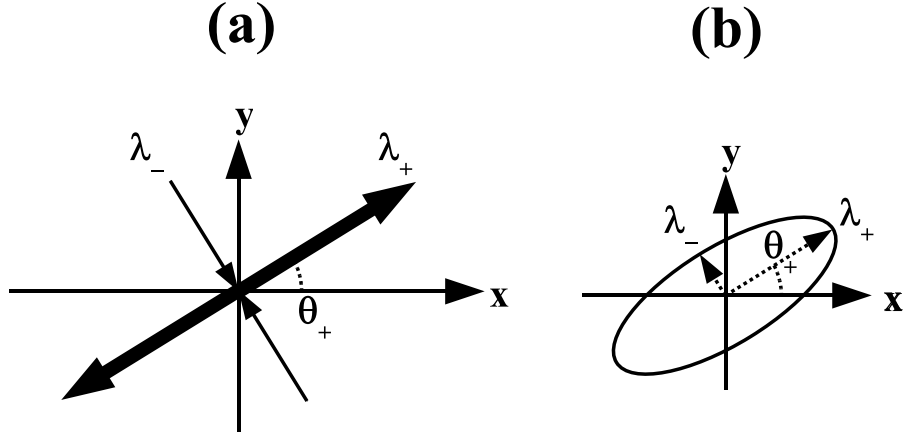}
\caption{\label{ReprTensors} Representation of symmetric tensors
(i.e. with orthogonal eigenvectors): (a) representation of a
tensor with two eigenvalues of different sign ($\lambda_- < 0 <
\lambda_+$). The thick (thin) line represents the direction and
magnitude of the positive (negative) eigenvalue; (b) elliptic
representation of a tensor with strictly positive eigenvalues ($0
< \lambda_- < \lambda_+$).}\end{center}
\end{figure}

As stated in Section \ref{Treatment}, we use the texture tensor as
a descriptor of bubble deformations. To be more quantitative, we
will use the statistical elastic strain tensor, defined as
(\cite{Aubouy2003}):\begin{equation}\label{StatDef} \bar{\bar{U}}
= \frac{1}{2} (\ln\bar{\bar{M}} - \ln\bar{\bar{M}}_0) ,
\end{equation}
where $\bar{\bar{M}}_0$ is a reference value, that we choose
isotropic: $\bar{\bar{M}}_0 = \lambda_0 \bar{\bar{I}}$. Here,
$\lambda_0$ is the average of the eigenvalues of the texture
tensors evaluated at the upstream and downstream extremities of
the observation field (left and right on Fig. \ref{Image}), where
the bubbles are less perturbed by the presence of the obstacle,
and $\bar{\bar{I}}$ is the 2D identity tensor. We use the
statistical elastic strain tensor because it quantifies the
elastic strain in foams, extending the classical elastic strain
(\cite{Landau1986}) to plastic flows. The trace of this tensor
quantifies the relative variation of area of the bubbles, which
remains lower than 10\% (\cite{Dollet2005a}); hence, in general,
this tensor will have a positive eigenvalue and a negative one. We
choose to represent such a tensor by two orthogonal lines, as
explained in Fig. \ref{ReprTensors}a. The positive (negative)
eigenvector represent the direction and amplitude of traction
(compression) of deformed bubbles compared to the reference state.
We also use this representation for the deformation rate which,
like the statistical elastic strain, is an almost traceless
tensor, since the flow remains weakly compressible. Here, the
positive (negative) eigenvector represent the direction and
amplitude of maximal elongation (compression) rate.

Finally, we represent the T1s by the two tensors
$\bar{\bar{T}}_+^{\mathrm{adim}}$ and
$\bar{\bar{T}}_-^{\mathrm{adim}}$ (Section \ref{Treatment}); these
two tensors are symmetric with strictly positive eigenvalues;
there is therefore suitably represented by ellipses (Fig.
\ref{ReprTensors}b) of parameterized equations:
$$ \left(%
\begin{array}{c}
  x_{\pm}(t) \\
  y(_{\pm}t) \\
\end{array}%
\right) = \bar{\bar{T}}_{\pm}^{\mathrm{adim}} \cdot \left(
\begin{array}{c}
  \cos t \\
  \sin t \\
\end{array}%
\right) = \left( \begin{array}{c}
  (T_{\pm}^{\mathrm{adim}})_{xx}\cos t + (T_{\pm}^{\mathrm{adim}})_{xy}\sin t \\
  (T_{\pm}^{\mathrm{adim}})_{xy}\cos t + (T_{\pm}^{\mathrm{adim}})_{yy}\sin t \\
\end{array}%
\right) . $$ In this case, the major axis is the preferred
direction for T1s to occur. We justify this (new) way to quantify
T1s in Section \ref{SectionT1}.

\section{Results} \label{Results}

In this section, we present the local measurements for a foam
flowing around an obstacle. We first focus on a reference case:
the flow of a monodisperse foam (bubble area: 16.0 mm$^2$, foam
thickness: 3.5 mm, bulk viscosity: 1.06 mm$^2$ s$^{-1}$, flow
rate: 176 ml min$^{-1}$) around a circular obstacle of diameter 30
mm. We compare averages and fluctuations of a local field to show
that the foam behaves like a continuous medium (section
\ref{AveragesFluctuations}). We then present a full study of the
reference case (section \ref{Reference}), and separate the
influence of each control parameter (section
\ref{ControlParameters}).

\subsection{Averages \emph{versus} fluctuations} \label{AveragesFluctuations}

Can we consider the foam as a continuous medium in our case? This
is not obvious \emph{a priori}, since the steady flow arises from
a balance between the load experienced by the bubbles passing
around the obstacle, and the discrete relaxations occurring during
T1 events (\cite{Langer1997}): locally-defined quantities like
elastic stress or statistical elastic strain fluctuate around an
average value. We consider here the influence of such fluctuations
and their correlations, since various studies have shown their
great importance, especially when T1 avalanches occur
(\cite{Debregeas2001,Kabla2003}), leading to strong stress drops
(\cite{Lauridsen2002,Pratt2003}).

\begin{figure}
\begin{center}
\includegraphics[width=12cm]{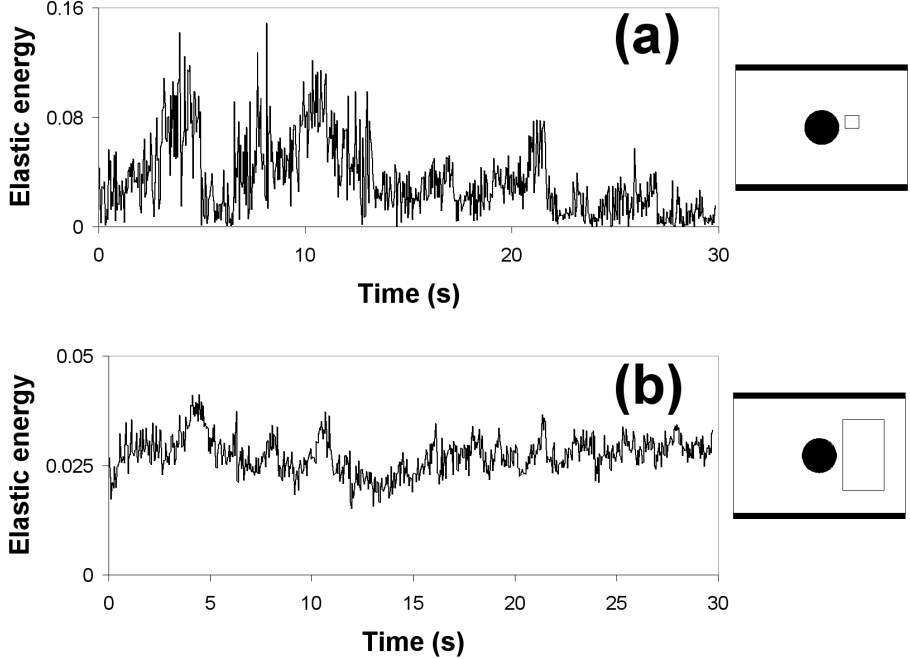}
\caption{\label{FluctPetitGrand} Temporal evolution of the
quantity $[\lambda_+(\bar{\bar{U}}) -
\lambda_-(\bar{\bar{U}})]^2$, which scales as the elastic energy.
(a) Results on a box of size $1.2\times 1.1$ cm$^2$ (sketched at
the right side). (b) Results on a box of size $3.6\times 6.2$
cm$^2$.}\end{center}
\end{figure}

\begin{figure}
\begin{center}
\includegraphics[width=12cm]{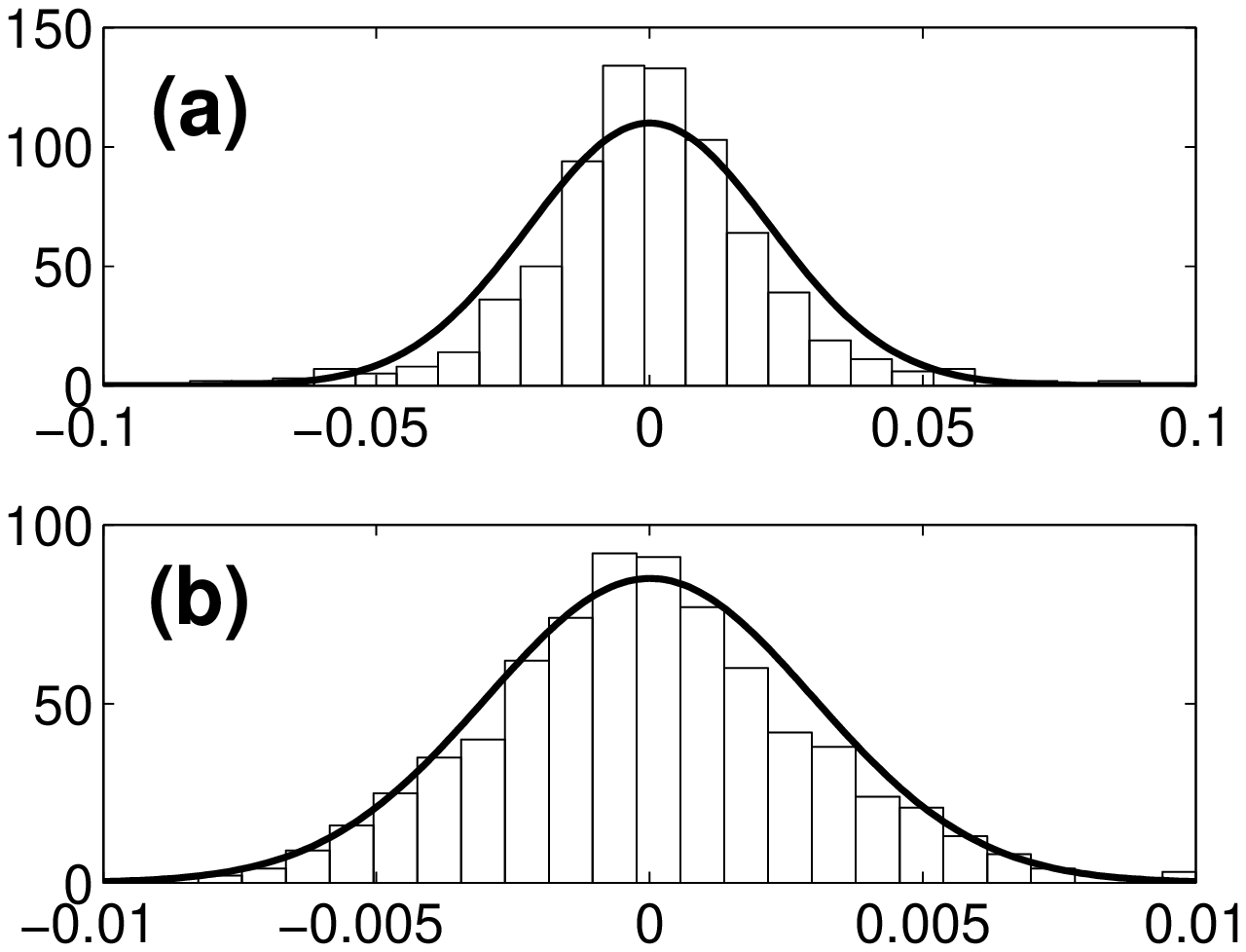}
\caption{\label{HistoPetitGrand} Histogram of the increments of
the the quantity $[\lambda_+(\bar{\bar{U}}) -
\lambda_-(\bar{\bar{U}})]^2$. (a) Results on a box of size
$1.2\times 1.1$ cm$^2$ (average: $3.7\times 10^{-5}$, standard
deviation: $2.2\times 10^{-2}$). (b) Results on a box of size
$3.6\times 6.2$ cm$^2$ (average: $-8.4\times 10^{-6}$, standard
deviation: $3.0\times 10^{-3}$). The curves superimposed are
Gaussian curves with the same mean and standard deviation as the
increments.}\end{center}
\end{figure}

To address this question, we analyse the temporal fluctuations of
a local quantity for our reference case, in the same spirit as
(\cite{Janiaud2005}). We have chosen the statistical elastic
strain tensor $\bar{\bar{U}}$, but the analysis would be similar
on other local quantities such as velocity or pressure. More
precisely, we have chosen to analyse one scalar quantity extracted
from $\bar{\bar{U}}$: the square of the difference between the two
eigenvalues $[\lambda_+(\bar{\bar{U}}) -
\lambda_-(\bar{\bar{U}})]^2$. This quantity scales as the elastic
energy associated to shear strain; hence, it is expected to
exhibit some huge drops if T1 avalanches occur, since they release
a lot of elastic energy. We have analysed the fluctuations in a
box close to the trailing side of the obstacle (Fig.
\ref{FluctPetitGrand}a, right), where fluctuations are expected to
be strong; we will see later that this is also a region where T1s
are frequent. The temporal variations of the bubble deformation is
reported in Fig. \ref{FluctPetitGrand}a. Qualitatively, we do not
observe a behavior dominated by T1 avalanches: such a behavior
would correspond to a succession of low increases (load) and quick
drops (relaxation) of the elastic energy. Quantitatively, we
analyse the increments of the bubble deformation between two
successive images, and report the histogram of the distribution of
these increments in Fig. \ref{HistoPetitGrand}a. This histogram is
well fitted by a Gaussian curve, characteristic of a white noise,
and we do not observe an asymmetric distribution with a lot of
small increases and a number of large decreases, which would
correspond to T1 avalanches.

However, the considered box is a small box, where only seven
bubbles in average are present at a given instant. One could thus
argue that fluctuations are dominated by advection, not by
possible T1 avalanches occurring at larger scale. We thus analyse
the fluctuations at a larger scale, choosing a box 18 times
bigger, in the wake of the obstacle (Fig. \ref{FluctPetitGrand}b,
right). The relative fluctuations are much smaller (Fig.
\ref{FluctPetitGrand}b), and the increments are here again well
fitted by a Gaussian curve (Fig. \ref{HistoPetitGrand}b).

We have performed this quantitative analysis of the fluctuations
only in our reference experiment, but we observed in the other
experiments analysed here that the fluctuations do not present
large-scale correlations; they are similar to a random, white
noise and play a negligible role at large scales. We will thus
only focus on coarse-grained average quantities, and treat the
foam as a continuous medium. The generality of such an approach is
discussed in Section \ref{DiscAveragesFluctuations}.

\subsection{Study of a reference case} \label{Reference}

\begin{figure}
\begin{center}
\includegraphics[width=12cm]{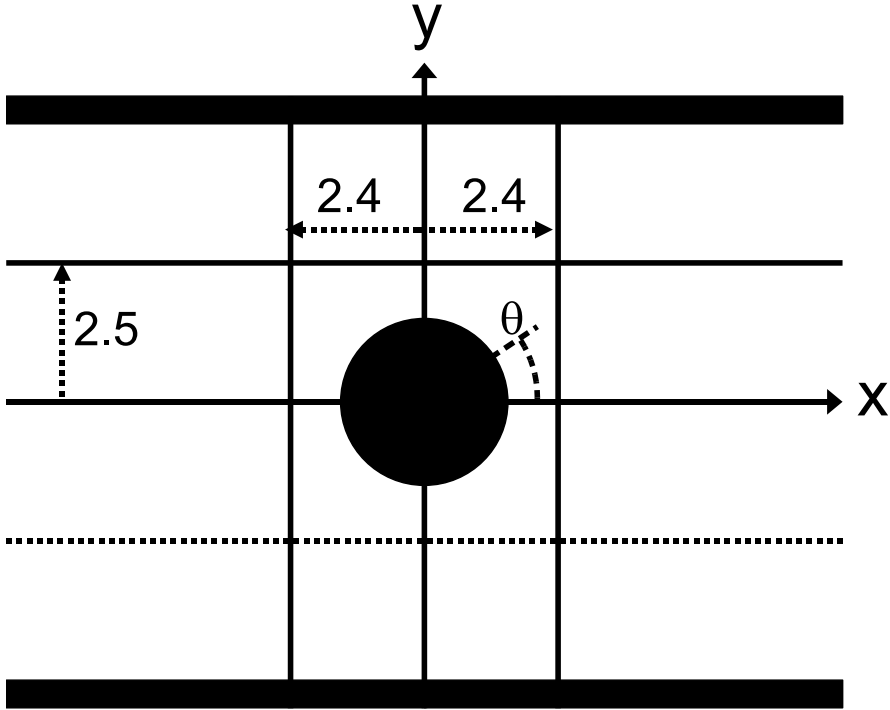}
\caption{\label{Lines} Sketch of the lines where the fields are
evaluated. The streamwise direction is $x$, the spanwise is $y$.
We choose five lines: the axis of symmetry of the flow, $y=0$, an
axis at the side of the obstacle, $|y|=2.5$ cm (the dashed axis
means that there are two such symmetric axes; the evaluated
quantities will be averaged on these both axes). Three axis are
perpendicular to the flow direction: $x=-2.4$, 0 and 2.4
cm.}\end{center}
\end{figure}

For each studied field, we proceed as follows: we first present a
map of the whole field, and we then study the variation of the
field components along various lines: two directed streamwise, one
on the axis of the obstacle and another aside, at 2.5 cm from the
axis of symmetry of the flow; and three directed spanwise: one
passing through the center of the obstacle, one upstream and the
symmetric downstream one, both lines being at 2.4 cm from the
middle axis (Fig. \ref{Lines}).

\subsubsection{Velocity} \label{RefVelocity}

\begin{figure}
\begin{center}
\includegraphics[width=12cm]{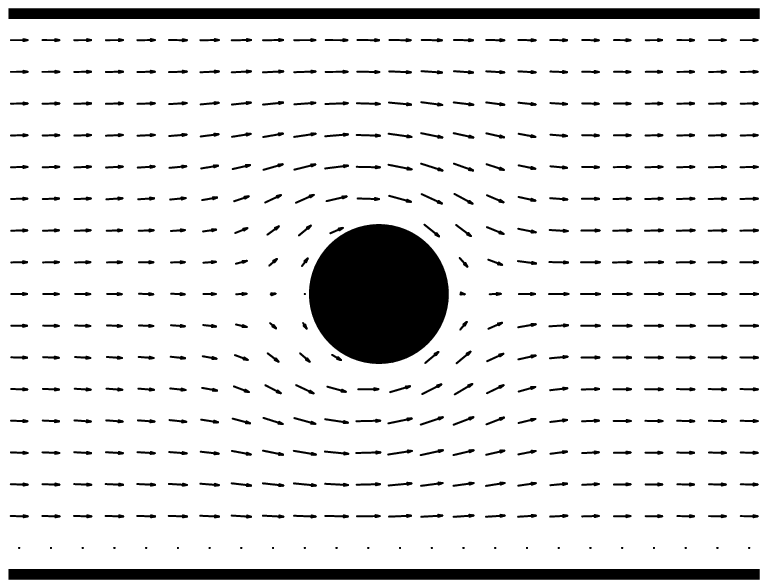}
\caption{\label{VelocityField} Velocity field around a circular
obstacle.}\end{center}
\end{figure}

The whole velocity field is presented in Fig. \ref{VelocityField}.
Qualitatively, the flow far from the obstacle is a plug flow, as
already observed for foam flows in narrow channels
(\cite{Cantat2004}). The obstacle imposes two symmetric stagnation
points, one upstream and one downstream, and the flow is
constricted, thus accelerated, on the sides of the obstacle.

\begin{figure}
\begin{center}
\includegraphics[width=12cm]{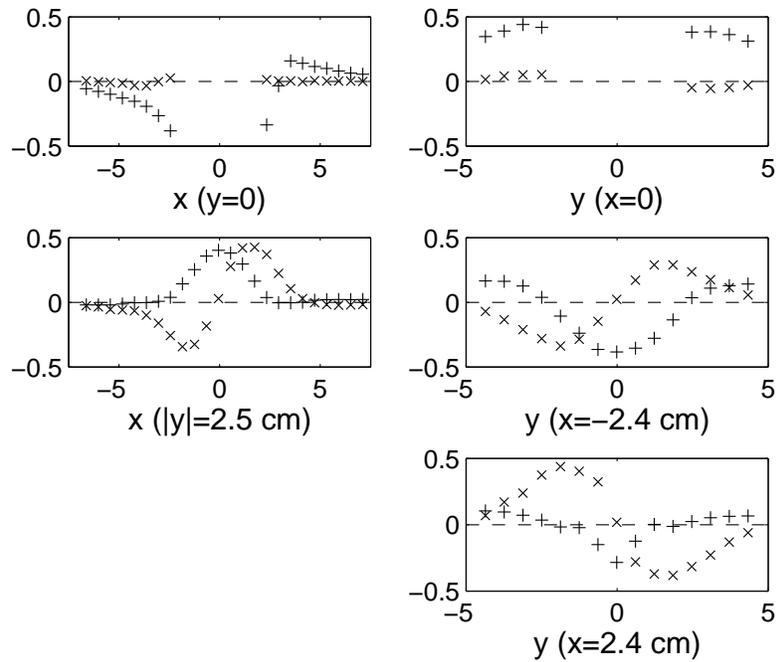}
\caption{\label{Velocity} Velocity components around a circular
obstacle: $(v_x-v_0)/v_0$ (+), and $v_y/v_0$ ($\times$). The
length unity is the centimeter.}\end{center}
\end{figure}

To study quantitatively the velocity, we divide it by the averaged
velocity $v_0$ obtained from the upstream and downstream
extremities of the observation region, where the flow is less
perturbed: $v_0$ is therefore the velocity of the plug flow. We
report the two components of $(\vec{v} - \vec{v}_0)/v_0$, which is
the dimensionless velocity deviation from the plug flow, in Fig.
\ref{Velocity}. This figure shows a striking feature: the velocity
is asymmetric up/downstream. More precisely, on the axis $y=0$,
the component $v_x$ shows an overshoot upstream, whereas it
decreases monotonically downstream. The asymmetry is also obvious
for the comparison of $v_x$ between the axis $x=-2.4$ and 2.4 cm:
the perturbation from the plug flow is higher upstream than
downstream for $v_x$, but lower for $v_y$. As expected, the $v_y$
component vanishes on the $y=0$ axis.

\subsubsection{Pressure} \label{RefPressure}

\begin{figure}
\begin{center}
\includegraphics[width=12cm]{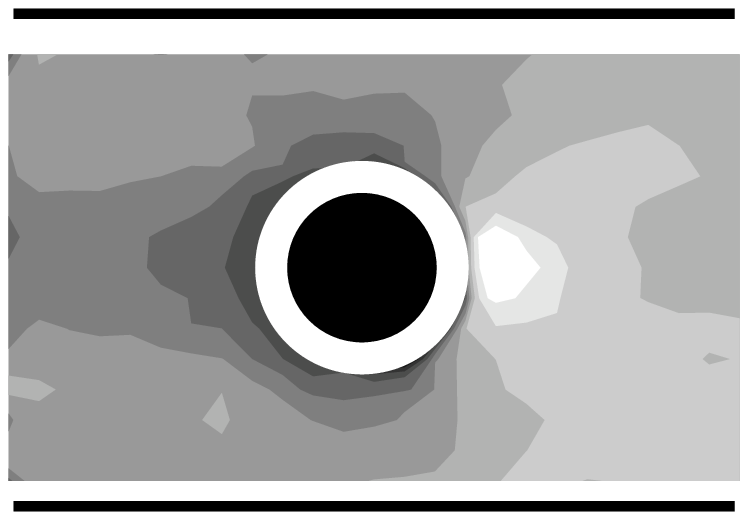}
\caption{\label{PressureField} Pressure field around a circular
obstacle. The higher the pressure, the darker the gray level. The
difference between two successive gray levels is 1.5 Pa,
corresponding to a relative variation in area of 2.5\%. The white
circle around the obstacle is the region where the bubble area,
hence the pressure, cannot be reliably evaluated.}\end{center}
\end{figure}

\begin{figure}
\begin{center}
\includegraphics[width=12cm]{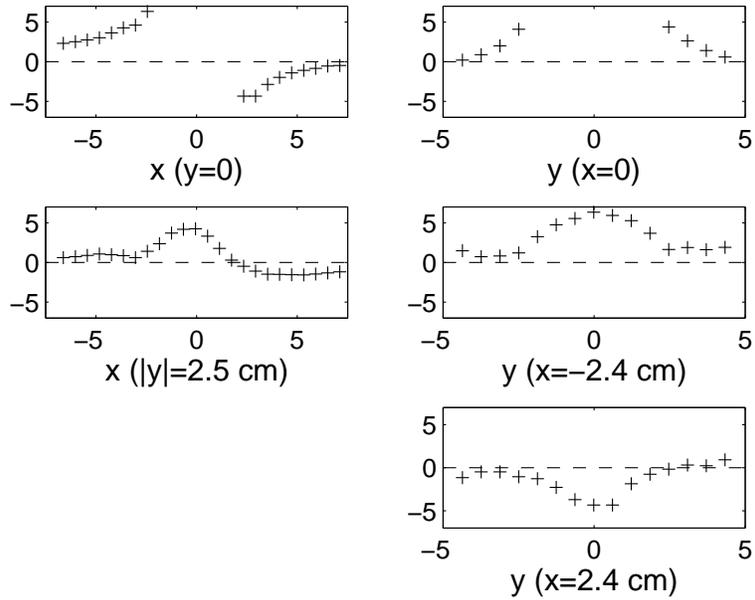}
\caption{\label{Pressure} Pressure evolution (in Pa) around a
circular obstacle. The pressure is obtained from the bubble area
from Eq. (\ref{P(A)_formula}); more precisely, we report here the
difference between the local pressure and an average one
corresponding to the average bubble area $A_0$.}\end{center}
\end{figure}

The whole pressure field is presented in Fig. \ref{PressureField}.
The pressure is maximal at the leading side of the obstacle, and
is minimal at its trailing side. We can also note that the
increase of pressure upstream is very progressive, extending
farther than the limits of the observation field. Fig.
\ref{Pressure} displays the evolution of the pressure along the
five axes of Fig. \ref{Lines}. We also observe an asymmetry
up/downstream; contrary to the velocity, this asymmetry is more
obvious on the side of the obstacle than on the axis $y=0$: the
pressure perturbation changes sign at $y=2$ cm. Note also that
along the axis $x=0$, the decrease of the perturbation in pressure
with the distance to the obstacle is quicker than the perturbation
in velocity.

\subsubsection{Statistical elastic strain}

\begin{figure}
\begin{center}
\includegraphics[width=12cm]{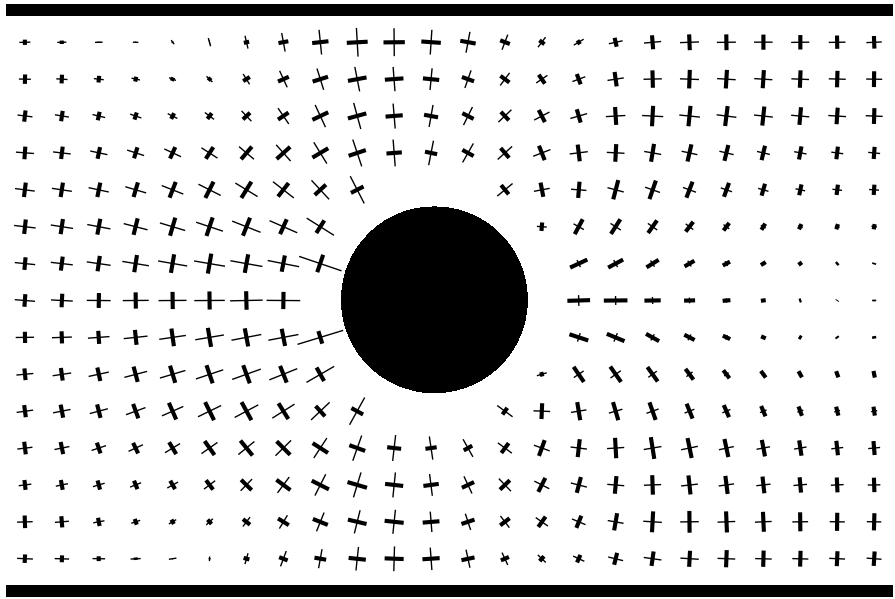}
\caption{\label{UField} Statistical elastic strain field around a
circular obstacle. The thick (thin) line is a direction of maximal
elongation (compression), see Fig.
\ref{ReprTensors}a.}\end{center}
\end{figure}

We now study the statistical elastic strain field, defined by Eq.
(\ref{StatDef}), to quantify the bubble deformation, as explained
in Section \ref{TensorialFields}. We display this field in Fig.
\ref{UField} using the representation explained in Fig.
\ref{ReprTensors}b. We note that the deformation is not negligible
for the bubbles entering the observation field: they are slightly
stretched in the spanwise direction. The bubbles are stretched in
the $x$ direction on the sides and in the wake of the obstacle,
and in the $y$ direction at the leading side of the obstacle and
on the sides of the wake.

\begin{figure}
\begin{center}
\includegraphics[width=12cm]{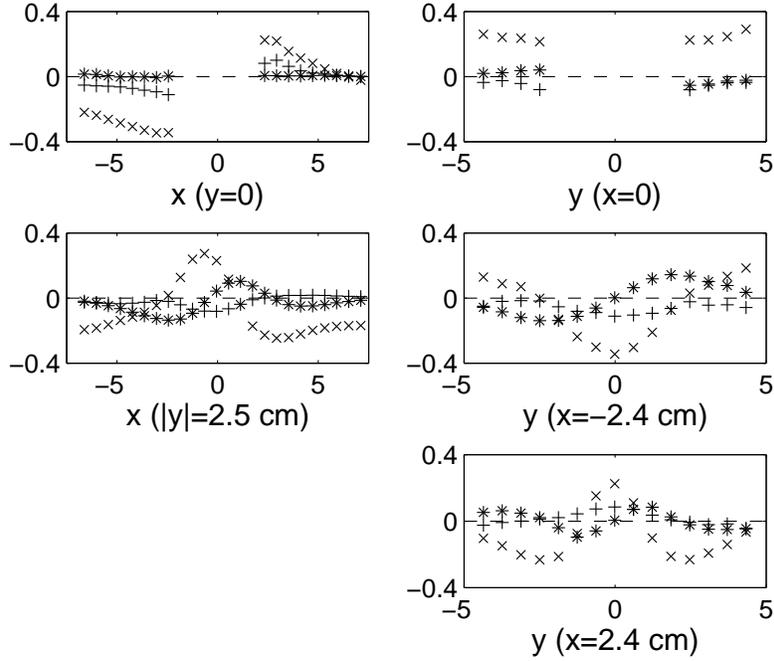}
\caption{\label{Deformation} Combinations $U_{xx}+U_{yy}$ (+),
$U_{xx}-U_{yy}$ ($\times$) and $U_{xy}$ ($*$) of the statistical
elastic strain field around a circular obstacle.}\end{center}
\end{figure}

The statistical elastic strain field $\bar{\bar{U}}$ is a
symmetric tensor, hence it has three independent components
$U_{xx}$, $U_{xy}$ and $U_{yy}$. Instead of these three
components, we have chosen to represent the combinations
$U_{xx}+U_{yy}$, $U_{xx}-U_{yy}$ and $U_{xy}$. The trace
$U_{xx}+U_{yy}$ gives access to the dilatation, whereas the
difference $U_{xx}-U_{yy}$ compares the bubble deformation in the
directions parallel and perpendicular to the flow, and the
$U_{xy}$ component indicates the deviation of the deformation from
the $x$ and $y$ directions. These combinations are plotted in Fig.
\ref{Deformation} along the five axis of Fig. \ref{Lines}. First,
the trace has a weak amplitude (its absolute value remains lower
than 0.1), and its evolution is strongly anti-correlated to the
pressure (Fig. \ref{Pressure}). The explanation of such a trend is
easy: when the pressure increases, the bubble area decreases as
explained in Section \ref{SectionPressure}. Hence, the length of
the vectors $\vec{r}$ linking centers of neighboring bubbles
decreases, and so does $U_{xx}+U_{yy} \approx \ln r/r_0$ after Eq.
(\ref{StatDef}). Second, we consider the parameter
$U_{xx}-U_{yy}$. It tends towards a negative value far from the
obstacle, which confirms the spanwise stretch observed on Fig.
\ref{UField}. This trend is observed both upstream and downstream,
hence we think that it is due to the longitudinal pressure
gradient due to the pressure drop along the channel. More
interestingly, the presence of the obstacle strongly modifies the
deformation of the bubbles: considering the downstream axis
($x=2.4$ cm), the bubbles are stretched streamwise in the wake,
and spanwise on the sides on the wake, the transition occurring at
$|y|=1.5$ cm. On the other hand, on the upstream axis ($x=-2.4$
cm) the bubbles are stretched spanwise close to the symmetry axis
of the flow, and streamwise on the sides, the transition occurring
at $|y|=2$ cm. Third, the $U_{xy}$ component on the sides of the
obstacle ($|y|=2.5$ cm) changes sign at two different points
($x=0$ and 2 cm), showing that the orientation of the maximal
deformation rotates of about $180^\circ$ during the passage around
the obstacle. Note also that this component is not strictly
reversed between upstream and downstream (comparison of the axis
$x=-2.4$ and $y=2.4$ cm).

\subsubsection{Velocity gradients}

\begin{figure}
\begin{center}
\includegraphics[width=12cm]{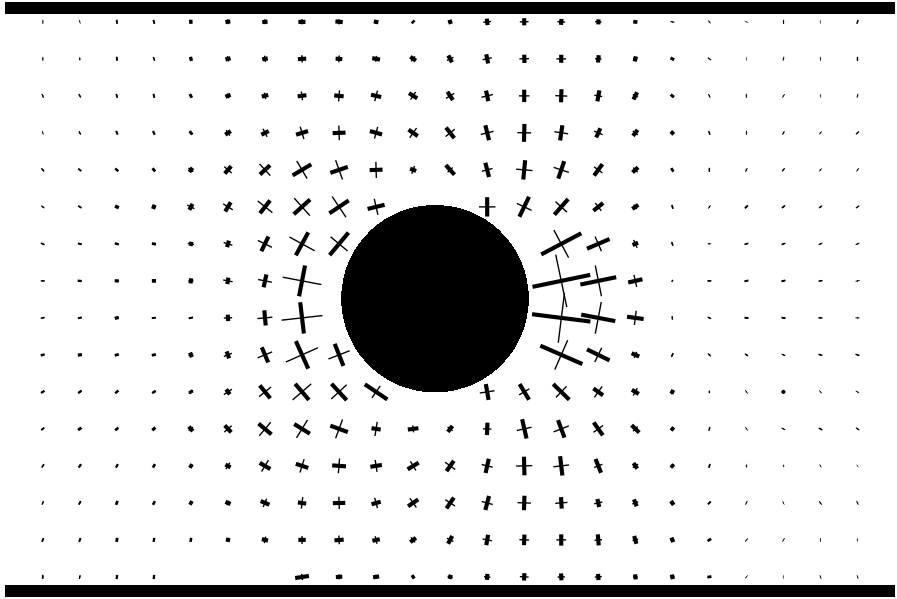}
\caption{\label{DeformationRate} Deformation rate field around a
circular obstacle. The thick (thin) line represents the maximal
elongation (compression) rate, see Fig.
\ref{ReprTensors}a.}\end{center}
\end{figure}

\begin{figure}
\begin{center}
\includegraphics[width=12cm]{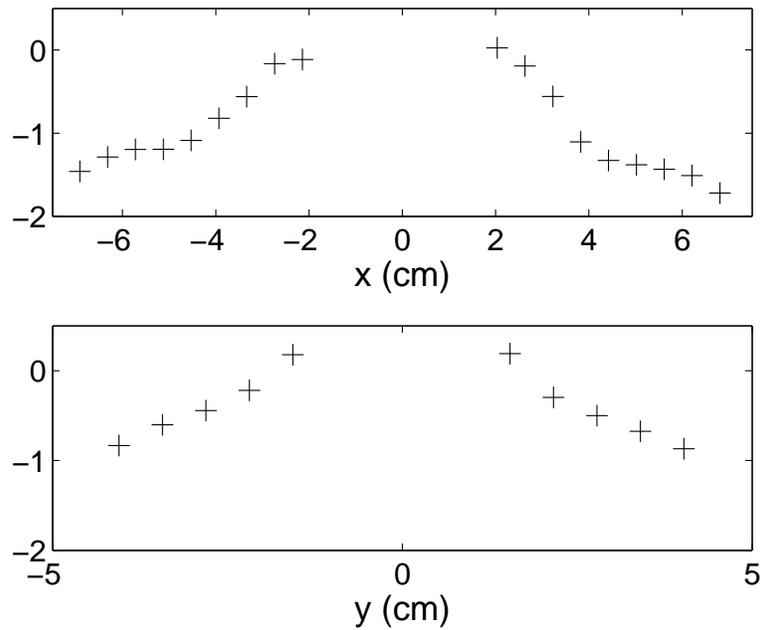}
\caption{\label{FonctDiss} Logarithm of the dissipation function
(expressed in s$^{-1}$) as a function of $x$ on the $y=0$ axis
(top), and as a function of $y$ on the $x=0$ axis
(bottom).}\end{center}
\end{figure}

We now turn to the velocity gradients. We first show the map of
the deformation rate in Fig. \ref{DeformationRate}. It confirms
that the 2D effective compressibility of the flow (see Section
\ref{SectionPressure}) remains weak, because the absolute values
maximal elongation and the maximal compression are very close at
any point. Furthermore, the amplitude of the deformation rate
decreases quickly with the distance to the obstacle, and seems to
become negligible upstream and downstream for a distance
comparable with the obstacle diameter. To investigate whether this
amplitude really vanishes at a finite distance from the obstacle,
as expected for a Bingham plastic in the same flow conditions
(\cite{Mitsoulis2004}), we consider the (scalar) dissipation
function (see Section \ref{TensorialFields}), and plot its
logarithm along the two symmetry axis $x=0$ and $y=0$ (Fig.
\ref{FonctDiss}). This plot reveals that the dissipation function
decreases with the distance of the obstacle, but does not vanish.
It is possible that it vanishes indeed farther from the obstacle,
but we have not investigated this possibility further.
Furthermore, the decrease is more complex than a power-law or
exponential decrease, and is quicker downstream than upstream.

\begin{figure}
\begin{center}
\includegraphics[width=12cm]{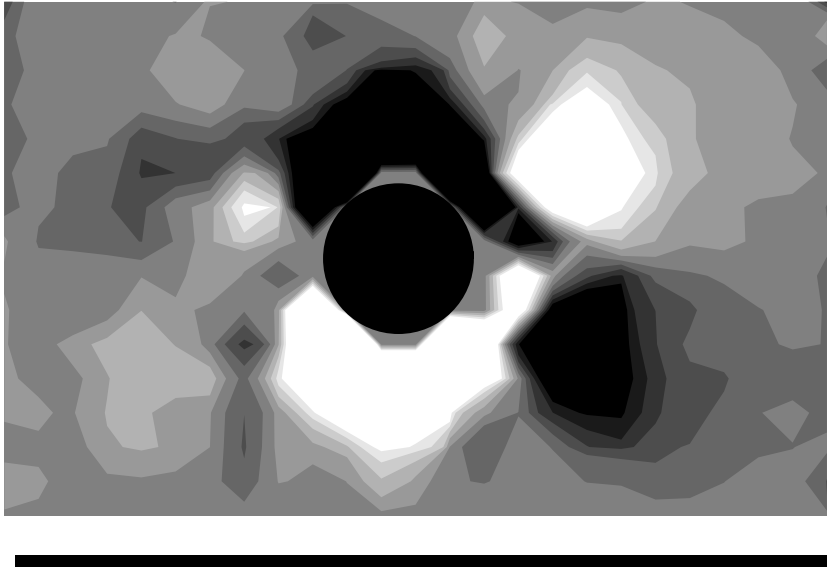}
\caption{\label{Vorticity} Vorticity field around a circular
obstacle. The light (dark) grey levels represent positive
(negative) vorticity.}\end{center}
\end{figure}

The map of the vorticity is presented in Fig. \ref{Vorticity}. The
vorticity exhibits significant variations, antisymmetric with
respect to the $y=0$ axis: in the $y>0$ half-channel, it is
negative on the side of the obstacle, and positive downstream
(there is also a little positive zone just before the obstacle).
The asymmetry up/downstream is once more obvious, as well as the
wake.

\subsubsection{T1 quantification} \label{SectionT1}

We first investigate the validity of our measurements of T1s. As
explained in Section \ref{Treatment}, the calculation of appearing
and disappearing edges is decoupled; therefore, we have to check
whether the number of these two kinds of events is the same, as
should be if we record correctly the T1s. Furthermore, we have
emphasized that our method may be sensitive to artifacts. We have
calculated the following quantity:
$$ \left. \sum_{\mathrm{every\, box\,} i} |f_+ - f_-|_i \right/ \sum_{\mathrm{every\, box\,} i} (f_+ +
f_-)_i = 7.0 \% , $$ which quantifies the relative uncertainty of
our method, acceptable despite the various sources of errors.

\begin{figure}
\begin{center}
\includegraphics[width=12cm]{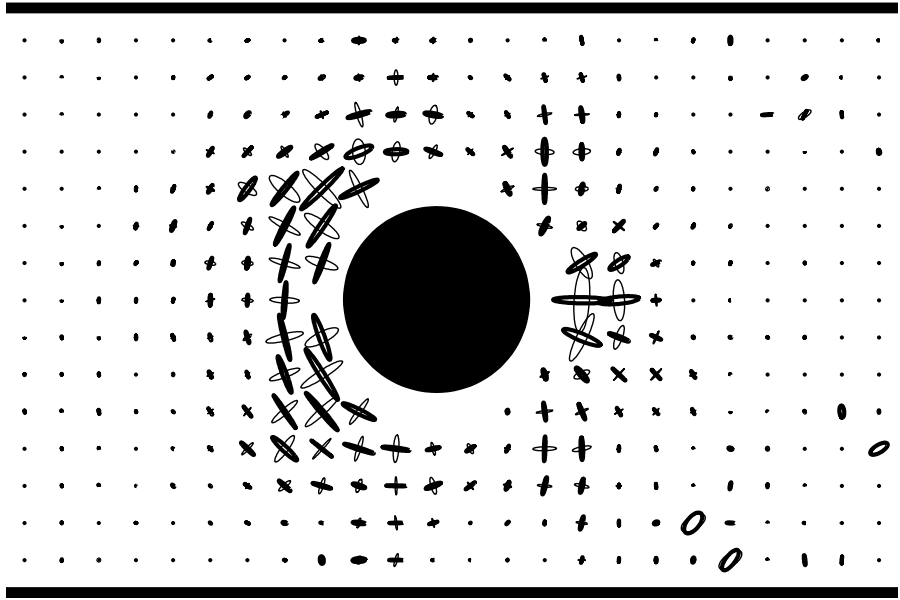}
\caption{\label{T1Ellipses} Elliptical representation of T1s: the
thick (thin) ellipses represent the tensor
$\bar{\bar{T}}_+^{\mathrm{adim}}$
($\bar{\bar{T}}_-^{\mathrm{adim}}$), defined in Section
\ref{Treatment}. The preferential direction of the T1s is obvious
in the adopted tensorial representation. Note the few remaining
artifacts (bottom right).}\end{center}
\end{figure}

\begin{figure}
\begin{center}
\includegraphics[width=12cm]{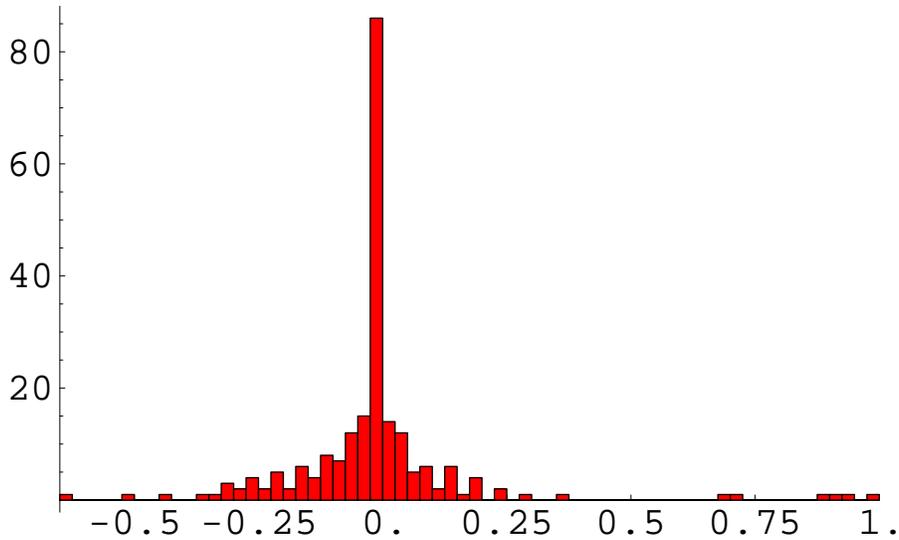}
\caption{\label{T1Orthogonal} Repartition histogram of the
quantity $\vec{x}_+ \cdot \vec{x}_-$ (see text for
definition).}\end{center}
\end{figure}

We now represent the map of T1s in Fig. \ref{T1Ellipses}, which
illustrates the advantages of the tensorial representation: not
only does it contain the number of T1s (proportional to the size
of the ellipses, as discussed in Section \ref{Treatment}), but
also their direction. The major axes of the two kinds of ellipses
are mainly orthogonal, which illustrates the fact that plastic
events release high stresses in one direction to the perpendicular
direction (\cite{Picard2004}). Quantitatively, denoting
$\vec{x}_+$ ($\vec{x}_+$) the unit vector of the major axis of the
ellipse representing $\bar{\bar{T}}_+^{\mathrm{adim}}$
($\bar{\bar{T}}_-^{\mathrm{adim}}$), we calculate for each box the
scalar product $\vec{x}_+ \cdot \vec{x}_-$, and report the
histogram of this quantity in Fig. \ref{T1Orthogonal}. This
histogram is actually strongly peaked around 0, confirming that
$\vec{x}_+$ and $\vec{x}_-$ are orthogonal. We also calculate the
average and standard deviation of the scalar product $\vec{x}_+
\cdot \vec{x}_-$, weighted by the number of T1s for each box:
$\langle \vec{x}_+ \cdot \vec{x}_- \rangle = 7.6\times 10^{-5}$
and $\delta(\vec{x}_+ \cdot \vec{x}_-) = 1.4\times 10^{-2} \ll 1$,
which proves the orthogonality of appearing and disappearing
edges.

Fig. \ref{T1Ellipses} shows that T1s are concentrated close to the
obstacle, but again with a significant asymmetry: upstream, the
T1s are more distributed and spread widely on the sides on the
obstacle, whereas downstream they are more localised in the wake.
Note also that the direction of T1s is correlated to the ones of
the statistical elastic strain (Fig. \ref{UField}) and of the
deformation rate (fig. \ref{DeformationRate}); such a correlation
is probably important to understand better the rheology of foams.

\begin{figure}
\begin{center}
\includegraphics[width=12cm]{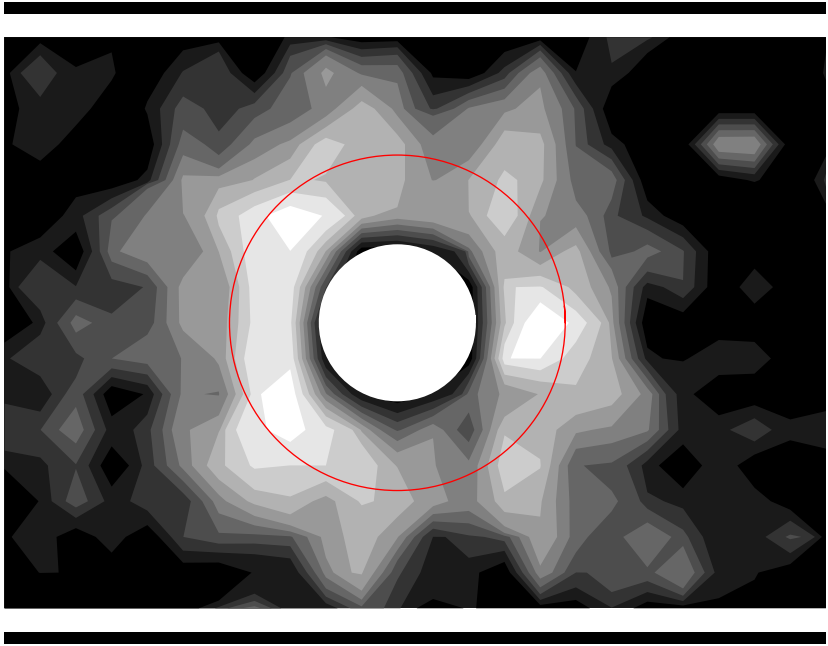}
\caption{\label{T1Norme2} Spatial distribution of the frequency of
T1s. The brighter the grey lever, the higher the frequency of T1s.
Note the attenuation of the artifacts in comparison with the Fig.
\ref{T1Ellipses}, as well as the marked asymmetry up/downstream.
The circle indicates the position chosen for the evaluation of
Fig. \ref{T1Polaire}.}\end{center}
\end{figure}

\begin{figure}
\begin{center}
\includegraphics[width=12cm]{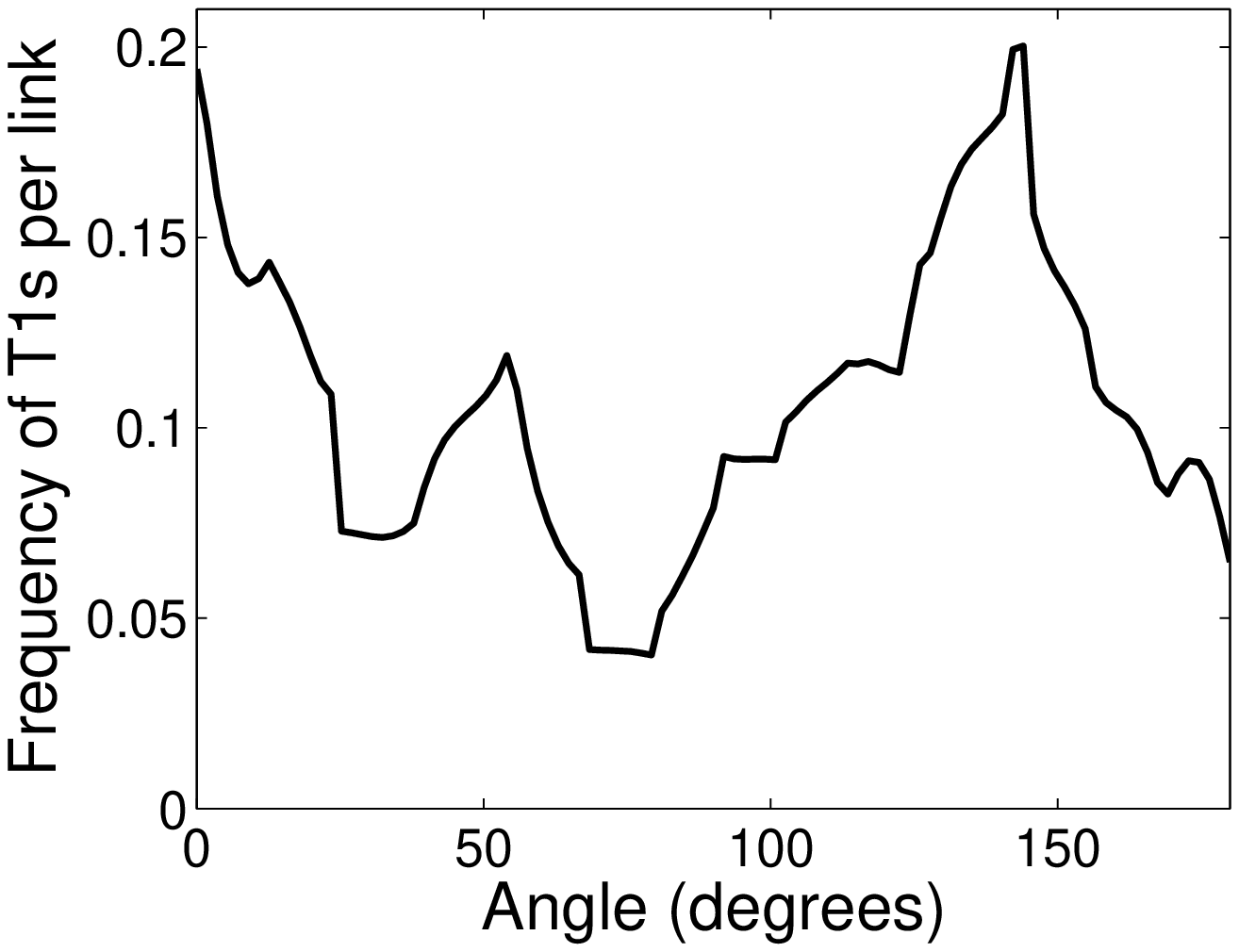}
\caption{\label{T1Polaire} Angular dependence (in degrees) of the
frequency of T1s per unit link, evaluated at 1.5 cm from the
boundary of the obstacle. The flow being symmetric with respect to
the axis $y=0$, the data have been averaged with the angles
between $-180$ and 0$^\circ$.}\end{center}
\end{figure}

To focus more on the spatial distribution of the frequency of T1s,
we now plot $\| \bar{\bar{T}}_+^{\mathrm{adim}} -
\bar{\bar{T}}_-^{\mathrm{adim}}\|/\sqrt{2} =
\sqrt{[(T_{+xx}^{\mathrm{adim}} - T_{-xx}^{\mathrm{adim}})^2 +
2(T_{+xy}^{\mathrm{adim}} - T_{-xy}^{\mathrm{adim}})^2 +
(T_{+yy}^{\mathrm{adim}} - T_{-yy}^{\mathrm{adim}})^2]/2}$. When
$\vec{x}_+ \cdot \vec{x}_- = 0$, which is in good approximation
true, this quantity equals $(f_+ + f_-)/2$; we thus identify it to
the frequency of T1s. This quantity has the advantage to reduce
strongly the remaining artifacts; its map is presented on Fig.
\ref{T1Norme2}. This map shows that the frequency of T1 presents
three maxima: one centered in the wake, and two symmetrically
off-centered downstream, at an angular position $|\theta| \simeq
3\pi/4$ (see Fig. \ref{Lines} for the definition of $\theta$). The
complex angular dependence of the T1 frequency is illustrated on
Fig. \ref{T1Polaire}. It shows that the off-centered downstream
maximum arises for an angle $\theta = 145^{\circ}$, and that
frequency of T1s is almost equal for this maximum and for the one
located in the wake. Fig. \ref{T1Polaire} shows also a secondary
maximum for $\theta = 55^{\circ}$.

\subsection{Influence of various control parameters} \label{ControlParameters}

In this section, we systematically study the dependence of the
fields describing the flow of foams around obstacles in the same
spirit as in \cite{Dollet2005a}: starting from the reference
experiment extensively studied in the previous section, we vary
only one control parameter, successively the flow rate (Section
\ref{InfluenceQ}), the bubble area (Section \ref{InfluenceA}), the
foam thickness (Section \ref{InfluenceE}) and the bulk viscosity
(Section \ref{InfluenceVisc}). To simplify the discussion, we only
study the evolution of three scalar quantities: the velocity
component $v_x$, the pressure $P$ and the quantity
$U_{xx}-U_{yy}$, along the axis of symmetry $y=0$. We end this
section by discussing the influence of the size and shape of the
obstacle (Section \ref{InfluenceObs}).

\subsubsection{Flow rate} \label{InfluenceQ}

\begin{figure}
\begin{center}
\includegraphics[width=12cm]{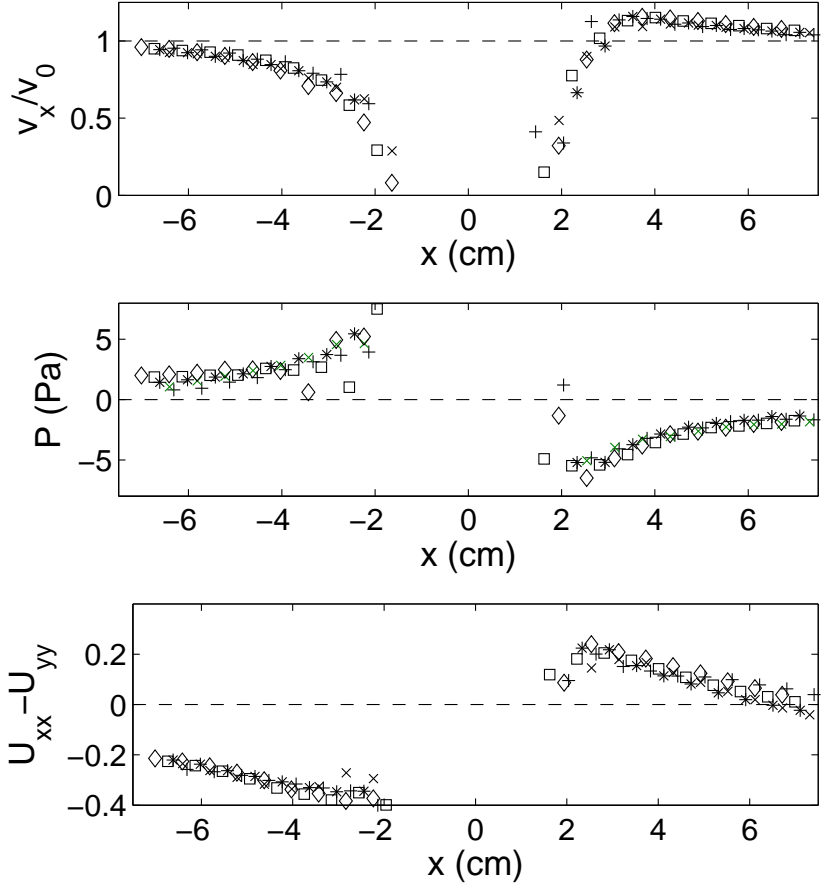}
\caption{\label{ComparisonQ} From top to bottom: plot of the
dimensionless velocity $v_x/v_0$, of the pressure, and of the
component $U_{xx}-U_{yy}$, as a function of $x$ along the axis
$y=0$, for the flow rates of 24 (+), 54 ($\times$), 176 ($*$), 293
($\Box$) and 515 ml min$^{-1}$ ($\diamond$).}\end{center}
\end{figure}

At given bubble area (16.0 mm$^2$), foam thickness (3.5 mm) and
bulk viscosity (1.06 mm$^2$ s$^{-1}$), we study five different
flow rates: 24, 54, 176, 293 and 515 ml min$^{-1}$ (corresponding
velocities $v_0$: 0.11, 0.26, 0.84, 1.40 and 2.45 cm s$^{-1}$). To
compare more easily the velocities, we consider the dimensionless
velocity $v_x/v_0$. We plot this quantity, as well as the pressure
and the component $U_{xx}-U_{yy}$, along the axis $y=0$, in Fig.
\ref{ComparisonQ}. Remarkably, all data points collapse on the
same master curve for the velocity, the pressure and the bubble
deformation, which proves that the qualitative features emphasized
in Section \ref{Reference} do not change on the studied range of
flow rate.

\subsubsection{Bubble area} \label{InfluenceA}

\begin{figure}
\begin{center}
\includegraphics[width=12cm]{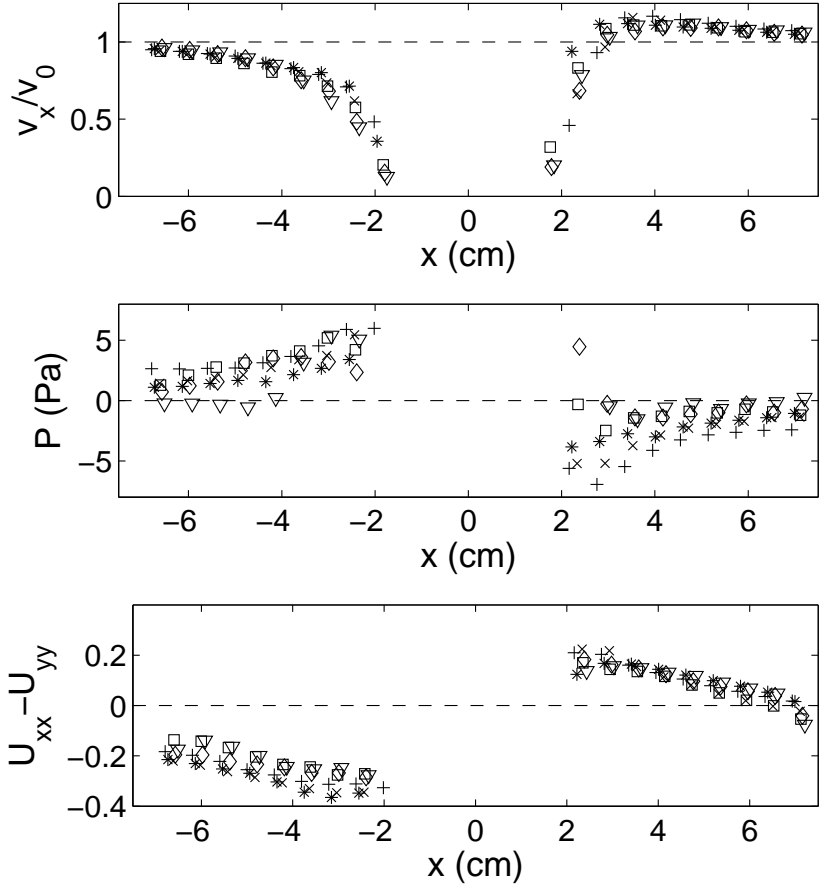}
\caption{\label{ComparisonA} From top to bottom: plot of the
dimensionless velocity $v_x/v_0$, of the pressure, and of the
component $U_{xx}-U_{yy}$, as a function of $x$ along the axis
$y=0$, for the bubble area of 12.1 (+), 16.0 ($\times$), 20.0
($*$), 25.7 ($\Box$), 31.7 ($\diamond$) and 39.3 mm$^2$
($\triangledown$).}\end{center}
\end{figure}

To study the influence of bubble area, the flow rate cannot be
strictly imposed, since it is slaved to target values of the other
control parameters. However, as shown in Section \ref{InfluenceQ},
it has no significant influence on the results. We study for each
bubble area: 12.1, 16.0, 20.0, 25.7, 31.7 and 39.3 mm$^2$, a flow
rate as close as possible to the one of the reference case,
respectively 160, 176, 166, 133, 150 and 169 ml min$^{-1}$. The
results are reported in Fig. \ref{ComparisonA}. They show that
neither the velocity field nor the bubble deformation depends
qualitatively on the bubble area. Only the pressure behavior in
the wake changes: for big enough bubbles, the pressure release at
the trailing side observed in Fig. \ref{Pressure} arises farther,
and can be preceded by a compression zone close to the obstacle.

\subsubsection{Foam thickness} \label{InfluenceE}

\begin{figure}
\begin{center}
\includegraphics[width=12cm]{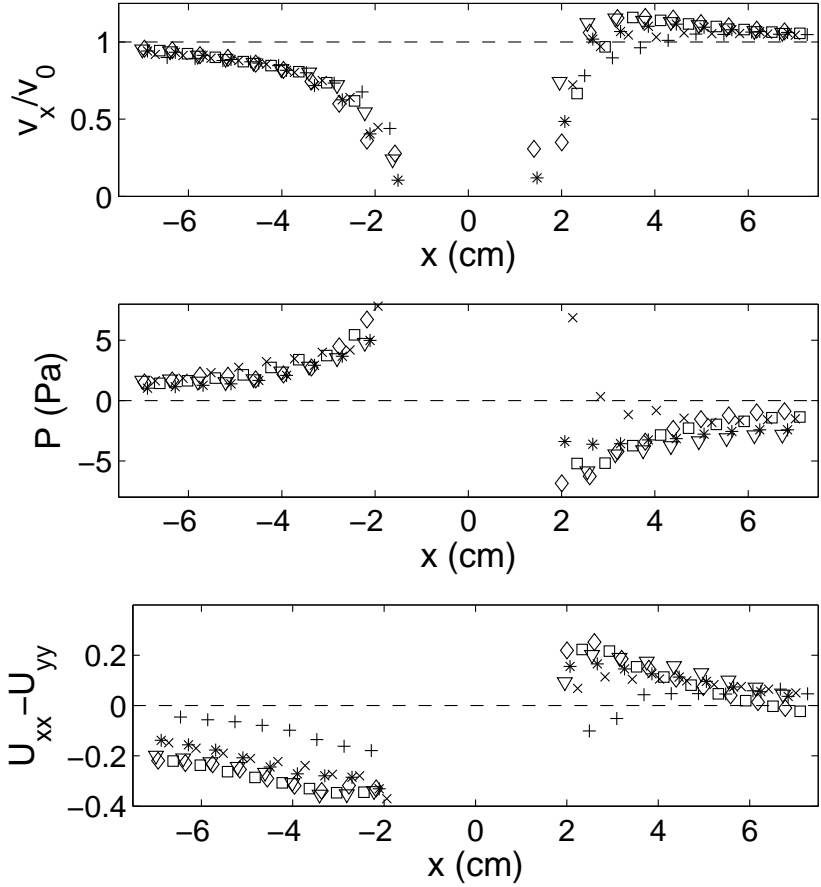}
\caption{\label{ComparisonE} From top to bottom: plot of the
dimensionless velocity $v_x/v_0$, of the pressure, and of the
component $U_{xx}-U_{yy}$, as a function of $x$ along the axis
$y=0$, for the foam thickness of 2.0 (+), 2.5 ($\times$), 3.0
($*$), 3.5 ($\Box$), 4.0 ($\diamond$) and 4.5 cm
($\triangledown$). Since the bubbles are decompacted, the pressure
cannot be calculated for the foam thickness of 2.0
mm.}\end{center}
\end{figure}

Various theoretical (\cite{Princen1983,Khan1986}) and experimental
(\cite{Princen1985,Mason1995,Mason1996,Saint-Jalmes1999}) studies
have shown that the fluid fraction plays a crucial role in the
foam rheology, but its influence on the local behavior of the foam
has so far not been studied in detail. In our case, the foam
thickness is a way to change the fluid fraction of the foam: the
thicker, the drier the foam. We study six different foam
thicknesses: 2.0, 2.5, 3.0, 3.5, 4.0 and 4.5 mm, at fixed bubble
area 16.0 mm$^2$. The corresponding fluid fractions are 8.5, 7.7,
7.4, 6.7, 6.4 and 5.8\% (see \cite{Raufaste} for the evaluation of
these fluid fractions). Since the cross section of the foam varies
proportionally to its thickness, we choose a mean velocity $v_0$
(see Section \ref{RefVelocity}) the closest possible to the one of
the reference case, respectively 0.67, 0.88, 0.89, 0.84, 0.74 and
0.56 cm s$^{-1}$ for the six thicknesses. Velocity, pressure and
bubble deformation are plotted in Fig. \ref{ComparisonE}. We
observe the following variations for the lowest foam thicknesses
(or highest fluid fractions): the asymmetry in the velocity is
weaker, there appears a compression zone in the wake close to the
obstacle (like for the biggest bubbles studied in Section
\ref{InfluenceA}), and the amplitude of the bubble deformation
decreases. Note that there is no significant variations for the
three highest thicknesses.

\subsubsection{Bulk viscosity} \label{InfluenceVisc}

\begin{figure}
\begin{center}
\includegraphics[width=12cm]{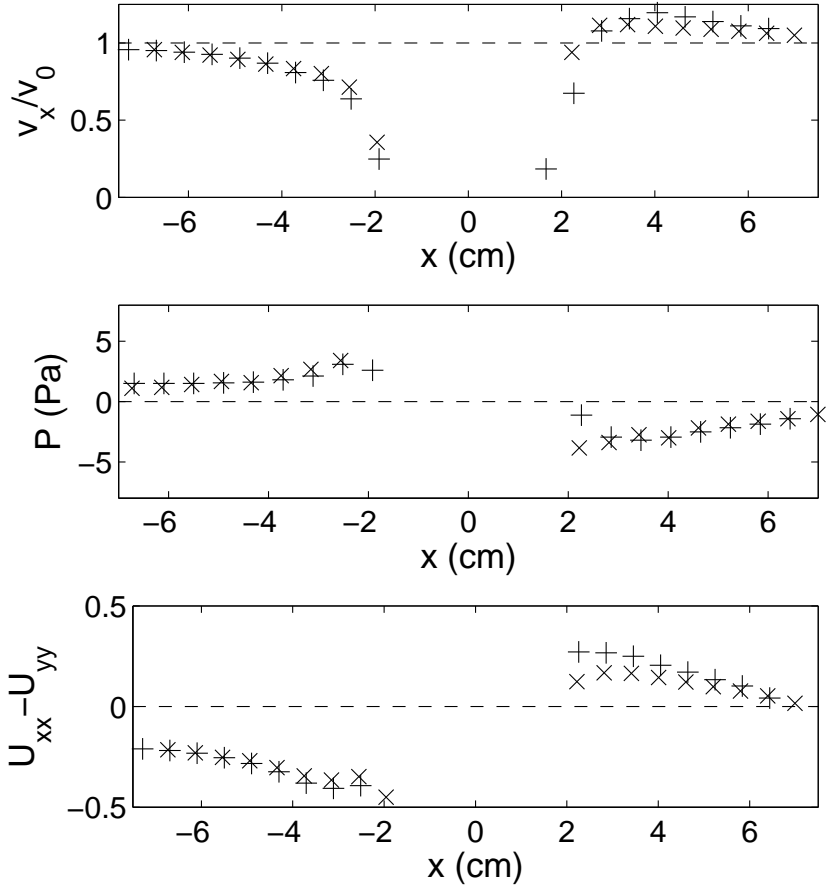}
\caption{\label{ComparisonV} From top to bottom: plot of the
dimensionless velocity $v_x/v_0$, of the pressure, and of the
component $U_{xx}-U_{yy}$, as a function of $x$ along the axis
$y=0$, for a bulk viscosity of 1.06 ($\times$), and 9.3 mm$^2$
s$^{-1}$ ($+$).}\end{center}
\end{figure}

We investigate now the influence of bulk viscosity. From the
various cases studied in \cite{Dollet2005a}, we only consider the
two extremes ones: a soap solution without added glycerol
(viscosity: 1.06 mm$^2$ s$^{-1}$), and another one with 50\% added
glycerol in mass (viscosity: 9.3 mm$^2$ s$^{-1}$). The bubble area
is 20.0 mm$^2$. The flow rates are 166 and 154 ml min$^{-1}$ for
the low and high viscosity cases. The results are presented on
Fig. \ref{ComparisonV}. They show that the bulk viscosity has no
significant effect, except in the wake close to the obstacle.

To summarize this study of the influence of the following control
parameters: flow rate, bubble area, foam thickness and bulk
viscosity, we emphasize that the main trends shown in a reference
case (Section \ref{Reference}) are robust, especially the
up/downstream asymmetry.

\subsubsection{Obstacle} \label{InfluenceObs}

\begin{figure}
\begin{center}
\includegraphics[width=12cm]{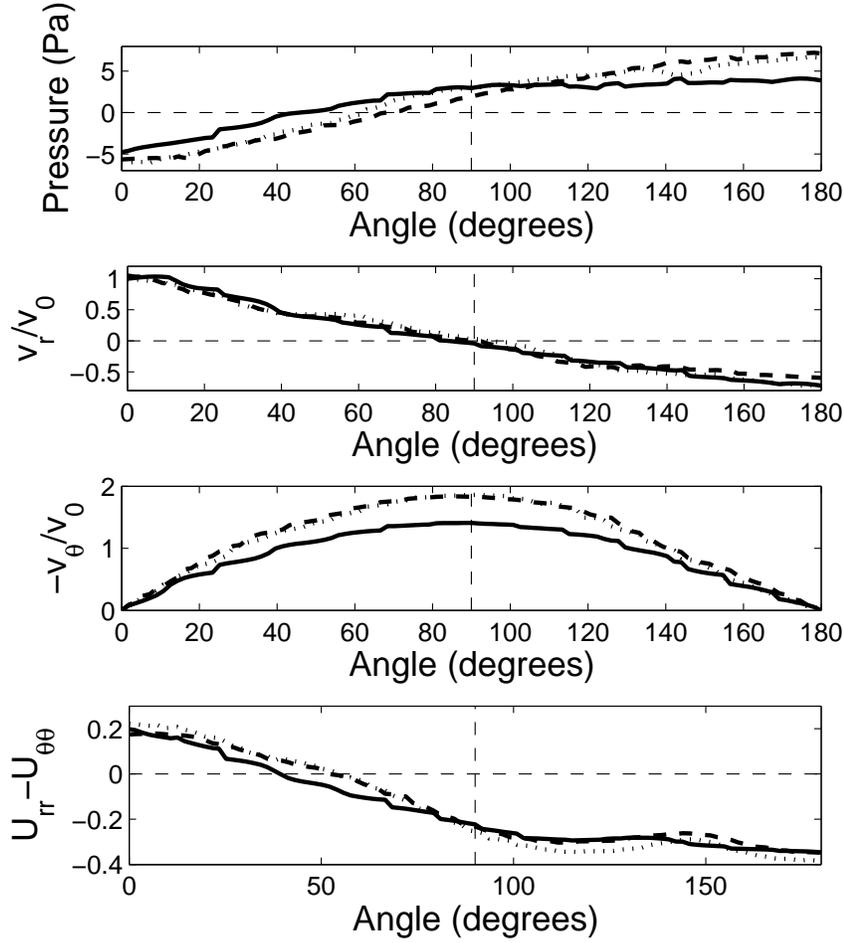}
\caption{\label{Obstacles} From top to bottom: plot of the
pressure, of the components $v_r/v_0$ and $-v_{\theta}/v_0$ of the
dimensionless velocity, and of the component $U_{rr} -
U_{\theta\theta}$ of the statistical elastic strain, for the
circles of diameter 30 mm (plain line) and 48 mm (long-dashed
line), and the cogwheel (short-dashed line) as a function of the
angle $\theta$, between 0 and 180$^\circ$. The flow being
symmetric with respect to the axis $y=0$, the data have been
averaged with the angles between $-180$ and 0$^\circ$.
}\end{center}
\end{figure}

The last control parameter that we have studied is the obstacle
itself. We have shown in previous studies that tuning the obstacle
geometry allows to show a variety of behaviors: streamlining for a
symmetric airfoil profile (\cite{Dollet2005a}) and anti-inertial
lift for a cambered one (\cite{Dollet2005b}), and combination of
drag, lift and torque for an elliptical obstacle
(\cite{DolletEllipse}). Here, we focus on simpler, circular
shapes, and compare the reference obstacle, the circle of diameter
30 mm, to a bigger circle, of diameter 48 mm, and a cogwheel of
diameter 43.5 mm with cogs of diameter 4 mm. The other control
parameters are the same that the reference experiments: foam
thickness of 3.5 mm, bulk viscosity of 1.06 mm$^2$ s$^{-1}$ and
bubble area of 16.0 mm$^2$. This area is suitable for bubbles to
be trapped into the cogs of the cogwheel, defining an effective
circular obstacle constituted by the cogwheel and the trapped
bubbles, of diameter 47.5 mm, comparable to the big circle. In
this paragraph, we thus study the influence of the size and of the
boundary of the obstacle. These obstacles sharing the circular
symmetry, we choose to study them in polar coordinates, plotting
the component $v_r$ and $-v_{\theta}$ of the velocity, the
pressure, and the deviatoric component of the statistical elastic
strain tensor $U_{rr} - U_{\theta\theta}$, as functions of
$\theta$ along a circle located at 1.5 cm from the obstacle
boundary (Fig. \ref{Obstacles}).

The data for the big circle and the cogwheel are very similar,
showing that the boundary conditions have little influence on the
behavior of the foam. The comparison between the two circles show
that whereas the radial component of the velocity is almost equal,
the amplitude of the orthoradial component is bigger for the
bigger circle. This is actually a consequence of the constriction
between the obstacles and the channel walls: more precisely, at
the angle $\theta = 90^\circ$, we have $-v_{\theta}/v_0 = v_y/v_0
= 1.83$ for the circle of diameter 48 mm, and $-v_{\theta}/v_0 =
1.41$ for the circle of diameter 30 mm, which is comparable to the
aspect ratio $h/(h-D)$, with $h=10$ cm the channel width and $D$
the obstacle diameter: this ratio equals 1.92 and 1.43 for these
two circles. Moreover, whereas the amplitude of variation of the
pressure is weaker for the small circle, the statistical elastic
strain component $U_{rr} - U_{\theta\theta}$ does not change much
between the three obstacles. It is negative and almost constant
for angles between 100 and 180$^\circ$ (Fig. \ref{Obstacles}),
which corresponds actually to an extended region where the yield
strain is reached; the component $U_{rr}-U_{\theta\theta}$ has a
markedly different behavior for angles between 0 and 100$^\circ$,
where it follows a monotonic, almost linear evolution, which is
the signature of an elastic-like transition to another yielded
region, situated in the wake of the obstacle. This is confirmed by
the polar dependence of the frequency of T1s (Fig.
\ref{T1Polaire}), which shows a high frequency in the yielded
regions for angles close to 0$^\circ$ and comprised between 100
and 180$^\circ$, whereas the frequency of T1s is lower in between.
A simple sketch of this behavior, with a yielded region on the
whole leading side of the obstacle and an elastic transition at
the trailing side to another yielded region in the wake, helps to
understand the fluid fraction dependence of the drag on circles
(\cite{Raufaste}) as well as the angular dependence of the drag,
lift and torque experienced by an ellipse (\cite{DolletEllipse}).

\section{Discussion}

\subsection{Liquid foam: localization or continuity?} \label{DiscAveragesFluctuations}

Our analysis of Section \ref{AveragesFluctuations}, showing that
the fluctuations are like a white noise, suggests that the foam
behaves as a continuous medium. We do not measure large-scale
correlations of plastic rearrangements. This is to compare to
flows involving pure shear, like experiments in Couette geometry
(\cite{Debregeas2001,Lauridsen2002,Pratt2003}). In this case, a
disordered and wet foam in a Hele-Shaw cell exhibits a strong
discontinuity in the form of a localised shear band
(\cite{Debregeas2001}), resulting from large-scale avalanches of
T1 (\cite{Kabla2003}), but a disordered and wet bubble raft shows
no such bands (\cite{Lauridsen2004}). This suggests that the
friction between the foam and the boundaries alters the continuity
of any quasi-2D foam. Indeed, a recent study has shown that at
given flow rate in the same Couette geometry, liquid pool foams
exhibit localization contrary to bubble rafts (\cite{Wang}).

However, our liquid pool foam does not exhibit localisation.
Actually, the flow around an obstacle is more complex than a pure
shear and screens the correlations of T1s, even though we have
very ordered foams. Qualitatively, we only saw large-scale
correlations with very wet and ordered foams, where dislocations
between rows of bubbles can occur on distances of several
centimeters. In the experiments presented here, it only occurs for
the smallest foam thickness.

\subsection{Discussion of the reference experiment}

\subsubsection{Velocity}

A salient feature of the flow of foam around an obstacle is the
asymmetry up/downstream of the velocity field. Such an asymmetry
is not captured by the models of Bingham plastics
(\cite{Roquet2003,Mitsoulis2004}) or Herschel-Bulkley fluids
(\cite{Beaulne1997}) usually invoked to simply model liquid foams.
It is an elastic effect, and the velocity overshoot in the wake of
the obstacle (Fig. \ref{Velocity}) has already been reported (and
termed negative wake) for a number of other elastic fluids (see
e.g. \cite{Dou2003,Kim2005}). However, a more quantitative
comparison to other viscoelastic fluids is difficult, because of
the boundary conditions: for a viscoelastic fluid, there is no
slip against a solid boundary, whereas at the macroscopic scale,
the foam slips against the walls and the obstacle (Fig.
\ref{VelocityField}). More precisely, a high velocity gradient is
concentrated in the thin liquid film between the solid boundaries
and the neighboring bubbles.

\subsubsection{Pressure}

We showed that the pressure in maximal at the upstream side of the
obstacle, and minimal at the downstream side (Section
\ref{RefPressure}), and that it does not depend significantly on
the flow rate (Section \ref{InfluenceQ}). This suggests that the
pressure is mainly of elastic origin, which is also corroborated
by the anti-inertial lift observed for an airfoil
(\cite{Dollet2005b}). More precisely, fitting the data for the
pressure along the axis $y=0$ (Fig. \ref{ComparisonQ}) by a
power-law yields the following dependence: $P(x) = (9.5\pm
0.7)x^{-1.13\pm 0.05}$ (expressed in Pa). The exponent is close to
$-1$, which is coherent with the stress distribution is an elastic
medium under a point-like force (\cite{Landau1986}).

%

\subsubsection{Coupling between statistical elastic strain, velocity gradients and T1s}

The asymmetric repartition of the T1s (Fig. \ref{T1Norme2}) is a
major result of this study, since the plastic flow of the foam
results from the superposition of many T1s. Since T1s more likely
occur for deformed bubbles, it is interesting to compare their
repartition with the map of statistical elastic strain (Fig.
\ref{UField}). Such a comparison reveals that the regions of
frequent T1s indeed correspond to high deformation, but the
correlation is not so simple: for instance, upstream from the
obstacle, the deformation is maximal at $x=0$ (Fig.
\ref{Deformation}), whereas the T1s occur more likely on the sides
of the obstacle, not at $x=0$ (Fig. \ref{T1Polaire}). This occurs
because in this region, the velocity gradients acts to increase
the preexistent deformation (Fig. \ref{DeformationRate}). There is
thus a strong coupling between the statistical elastic strain, the
velocity gradients and the T1 repartition, which is analysed in
more detail elsewhere (\cite{Graner}).

\subsection{Influence of the control parameters and comparison with force measurements}

We now compare qualitatively our results to the force measurements
realized in the same conditions in \cite{Dollet2005a}. To
summarize, this study showed that the drag exerted by a flowing
foam on a circular obstacle scales as:
\begin{equation}\label{ScalingForce}
  F = F_0 + \mathrm{cste}\times\eta^{0.77}v_0,
\end{equation}
 with $\eta$ the bulk viscosity, and $F_0$ a decreasing function of the bubble area. The independence
of the pressure and the bubble deformation on the flow rate is in
qualitative agreement with the linear increase of the force
exerted by the flowing foam on the obstacle (\cite{Dollet2005a}),
if we assume that the velocity-dependent contribution to the force
is mainly due to the viscous friction in the liquid films between
the obstacle and the surrounding bubbles. Moreover, the exponent
for the bulk viscosity in (\ref{ScalingForce}) is close to 1,
which is compatible with the weak influence of this parameter
(Fig. \ref{ComparisonV}). Concerning the foam thickness, Fig.
\ref{ComparisonE} shows that the up/downstream differences
increases with the foam thickness, hence decreases with the fluid
fraction. This is compatible with the decrease of the drag with
the fluid fraction reported in \cite{Raufaste}. The role of the
bubble area is more complex: we showed in Section \ref{InfluenceA}
that this parameter does not influence much the bubble
deformation, but that the up/downstream difference in pressure
decreases with increasing bubble area. The bubble deformation and
pressure being the two contributions to the yield drag $F_0$ in
(\ref{ScalingForce}), this qualitatively agrees with the fact that
$F_0$ decreases with the bubble area. However, we do not
understand yet why the behavior of the pressure and deformation
differs so markedly. The last studied control parameter, the
obstacle itself, shows that the boundary conditions at the
obstacle plays no significant role. The size of the obstacle has
two influences: the bigger the circle, the higher the velocity at
its sides, due to the imposed constriction, and the higher the
amplitude of variation of the pressure. This is compatible with
the measured values of the drag coefficient (ratio of drag and
obstacle radius) reported in \cite{Dollet2005a} for these three
obstacles, equal for the cogwheel and the big circle, and slightly
lower for the small circle.

\section{Conclusions}

We performed a systematic local description of the flow of foam
around a circular obstacles, quantifying the elasticity by the
statistical elastic strain tensor and the pressure field, the
plasticity by a tensorial descriptor of bubble rearrangements, and
the flow by the velocity field and its gradients. We demonstrate
in Section \ref{AveragesFluctuations} that such a continuous
approach is justified and suitable in this case. The study of a
reference experiment (Section \ref{Reference}) shows a marked
asymmetry up/downstream: the velocity exhibits an overshoot in the
wake, and the bubble rearrangements spread more at the leading
side than in the wake of the obstacle. This reveals the complex
behavior of the foam, dictated by the coupling between elasticity
(bubble deformations), plasticity (bubble rearrangements) and flow
(velocity gradients).

We showed in Section \ref{InfluenceQ} that the rescaled velocity
$\vec{v}/v_0$, the pressure and the bubble deformation do not
depend significantly on the flow rate, in the studied range.
Hence, the yield and the dynamic contributions in foam rheology
seem to be decoupled, which justifies recent approaches to model
foam flows (and more generally, complex fluid flows) as Bingham
plastics with an added elastic term independent of the shear rate
(\cite{Takeshi2005,Weaire}). However, such scalar approaches are
not sufficient in our context, where the complex flow around an
obstacle appeals for a tensorial description. It would be
interesting to perform the same experiments at higher flow rate,
to determine when this decoupling between elastic and dynamic
contribution breaks down, and to investigate whether phenomena
such as elastic turbulence (\cite{Groisman2000}) could then occur
in the wake of the obstacle.

The detailed results shown in Section \ref{Results}, combined with
the associated force measurements reported in \cite{Dollet2005a},
severely constrain the rheological models adapted to describe
liquid foams. The foam flowing around an obstacle intrinsically
exhibits yield stress and dissipation (yield drag on the obstacle
(\cite{deBruyn2004,Dollet2005a})), and elastic stress
(up/downstream asymmetry); all these effects can only be captured
by a whole elastoviscoplatic model.


\begin{thebibliography}{99}

\bibitem[Asipauskas \emph{et al.} (2003)]{Asipauskas2003} \textsc{M. Asipauskas, M. Aubouy, J. A. Glazier, F.
Graner \& Y. Jiang} 2003 A texture tensor to quantify
deformations: The example of two-dimensional flowing foams.
\emph{Granular Matt.} \textbf{5}, 71--74.

\bibitem[Aubouy \emph{et al.} (2003)]{Aubouy2003} \textsc{M. Aubouy, Y. Jiang, J. A. Glazier \& F. Graner} 2003
A texture tensor to quantify deformations. \emph{Granular Matt.}
\textbf{5}, 67--70.

\bibitem[Batchelor (1970)]{Batchelor1970} \textsc{G. K. Batchelor} 1970
The stress system in a suspension of force-free particles.
\emph{J. Fluid Mech.} \textbf{41}, 545--570.

\bibitem[Beaulne \& Mitsoulis (1997)]{Beaulne1997} \textsc{M. Beaulne \& E. Mitsoulis} 1997
Creeping motion of a sphere in tubes filled with Herschel--Bulkley
fluids \emph{J. Non-Newtonian Fluid Mech.} \textbf{72}, 55--71.

\bibitem[Cantat \& Delannay (2003)]{Cantat2003} \textsc{I. Cantat \& R. Delannay}
2003 Dynamical transition induced by large bubbles in
two-dimensional foam flows.
 \emph{Phys. Rev. E} \textbf{67}, 031501.

\bibitem[Cantat, Kern \& Delannay (2004)]{Cantat2004} \textsc{I. Cantat, N. Kern \& R.
Delannay} 2004 Dissipation in foam flowing through narrow
channels.
 \emph{Europhys. Lett.} \textbf{65}, 726--732.

\bibitem[Cantat \& Delannay (2005)]{Cantat2005} \textsc{I. Cantat \& R. Delannay} 2005
Dissipative flows of 2D foam. \emph{Eur. Phys. J. E} \textbf{18},
55--67.

\bibitem[Cantat \& Pitois (2005)]{Cantat2005b} \textsc{I. Cantat \& O. Pitois} 2005
Mechanical probing of liquid foam ageing. \emph{J. Phys. Condens.
Matt.} \textbf{17}, S3455--S3461.

\bibitem[Courty \emph{et al.} (2003)]{Courty2003} \textsc{S. Courty, B. Dollet, F. Elias, P. Heinig \& F. Graner}
2003 Two-dimensional shear modulus of a Langmuir foam.
\emph{Europhys. Lett.} \textbf{64}, 709--715.

\bibitem[Cox \emph{et al.} (2000)]{Alonso2000} \textsc{S.
J. Cox, M. D. Alonso, S. Hutzler, D. Weaire} 2000 The Stokes
experiment in a foam, in \emph{Proceedings of the 3rd
Euroconference on Foams, Emulsions and their Applications}, P. L.
J. Zitha, J. Banhard, P. L. M. M. Verbist Eds., MIT Verlag,
Bremen, 282--289 (2000).

\bibitem[Cox, Vaz \& Weaire (2003)]{Cox2003}
\textsc{S. Cox,  M. F.   Vaz \& D. Weaire} 2003.  \emph{Euro.
Phys. J. E} \textbf{11}, 29--35.

\bibitem[de Bruyn (2004)]{deBruyn2004} \textsc{J. R. de Bruyn}
2004 Transient and steady-state drag in foam. \emph{Rheol. Acta}
\textbf{44}, 150--159.

\bibitem[Debr\'egeas, Tabuteau \& di Meglio (2001)]{Debregeas2001} \textsc{G. Debr\'egeas, H. Tabuteau \& J.-M. di
Meglio} 2001 Deformation and flow of a two-dimensional foam under
continuous shear. \emph{Phys. Rev. Lett.} \textbf{87}, 178305.

\bibitem[Denkov \emph{et al.} (2005)]{Denkov2005} \textsc{N. D. Denkov, V. Subramanian, D. Gurovich \& A.
Lips} 2005 Wall slip and viscous dissipation in sheared foams:
Effect of surface mobility. \emph{Coll. Surf. A} \textbf{263},
129--145.

\bibitem[Derjaguin (1933)]{Derjaguin1933} \textsc{B. Derjaguin} 1933 Die elastischen Eigenschaften der
Sch\"aume. \emph{Kolloid Z.} \textbf{64}, 1--6.

\bibitem[Dollet \emph{et al.} (2005a)]{Dollet2005a} \textsc{B. Dollet, F. Elias, C. Quilliet, C. Raufaste, M.
Aubouy \& F. Graner} 2005a Two-dimensional flow of foam around an
obstacle: Force measurements. \emph{Phys. Rev. E} \textbf{71},
031403.

\bibitem[Dollet \emph{et al.} (2005b)]{Dollet2005c} \textsc{B. Dollet, F. Elias, C. Quilliet, A. Huillier, M.
Aubouy \& F. Graner} 2005b Two-dimensional flows of foam: Drag
exerted on circular obstacles and dissipation. \emph{Colloids
Surf. A} \textbf{263}, 101--110.

\bibitem[Dollet, Aubouy \& Graner (2005)]{Dollet2005b} \textsc{B. Dollet, M. Aubouy \& F. Graner} 2005
Anti-inertial lift in foams: A signature of the elasticity of
complex fluids. \emph{Phys. Rev. Lett.} \textbf{95}, 168303.

\bibitem[Dollet, Durth \& Graner (2006)]{DolletEllipse} \textsc{B. Dollet, M. Durth \& F. Graner} 2006
Flow of foam past an elliptical obstacle, to appear in \emph{Phys.
Rev. E}, \texttt{arXiv:cond-mat/0601100}.

\bibitem[Dou \& Phan-Thien (2003)]{Dou2003} \textsc{H. S. Dou \& N. Phan-Thien} 2003
Negative wake in the uniform flow past a cylinder. \emph{Rheol.
Acta} \textbf{42}, 383--409.

\bibitem[Groisman \& Steinberg (2000)]{Groisman2000} \textsc{A. Groisman \& V. Steinberg} 2000
Elastic turbulence in a polymer solution flow \emph{Nature}
\textbf{405}, 53--55.

\bibitem[Guyon, Hulin \& Petit (2001)]{Guyon2001} \textsc{\'E. Guyon, J.-P. Hulin \& L. Petit} 2001 \emph{Hydrodynamique
physique}, EDP Sciences/CNRS \'Editions, Paris.

\bibitem[H\"ohler \& Cohen-Addad (2005)]{Hohler2005} \textsc{R. H\"ohler \& S. Cohen-Addad} 2005
Rheology of liquid foams \emph{J. Phys. Condens. Matter}
\textbf{17}, R1041--R1069.

\bibitem[Janiaud \& Graner (2005)]{Janiaud2005} \textsc{\'E. Janiaud \& F. Graner} 2005
Foam in a two-dimensional Couette shear: A local measurement of
bubble deformation. \emph{J. Fluid Mech.} \textbf{532}, 243--267.

\bibitem[Kabla \& Debr\'egeas (2003)]{Kabla2003} \textsc{A. Kabla \& G. Debr\'egeas}
2003 Local stress relaxation and shear-banding in a dry foam under
shear. \emph{Phys. Rev. Lett.} \textbf{90}, 258303.

\bibitem[Kern \emph{et al.} (2004)]{Kern2004} \textsc{N. Kern, D. Weaire, A. Martin, S. Hutzler \& S. J.
Cox} 2004 Two-dimensional viscous froth model for foam dynamics,
\emph{Phys. Rev. E} \textbf{70}, 041411.

\bibitem[Khan \& Armstrong (1986)]{Khan1986} \textsc{S. A. Khan \& R. C. Armstrong} 1986 Rheology of foams. I. Theory for dry
foams. \emph{J. Non-Newtonian Fluid Mech.} \textbf{22}, 1--22.

\bibitem[Khan \& Prud'homme (1996)]{Khan1996} \textsc{S. A. Khan \& R. Prud'homme} 1996 \emph{Foams},
Dekker, New York.

\bibitem[Kim \emph{et al.} (2005)]{Kim2005} \textsc{J. M. Kim, C. Kim, C. Chung, K. H. Ahn \& S. J.
Lee} 2005 Negative wake generation of FENE--CR fluids in uniform
and Poiseuille flow past a cylinder.
 \emph{Rheol. Acta} \textbf{44}, 600--613.

\bibitem[Landau \& Lifshitz (1986)]{Landau1986} \textsc{L. D. Landau \& E. M. Lifshitz} 1986 \emph{Theory of
elasticity}, 3rd edition, Reed, Oxford.

\bibitem[Langer \& Liu (1997)]{Langer1997} \textsc{S. A. Langer \& A. J. Liu} 1997
Effect of random packing on stress relaxation in foam. \emph{J.
Phys. Chem. B} \textbf{101}, 8667--8671.

\bibitem[Larson (1999)]{Larson1999} \textsc{R. G. Larson} 1999 \emph{The structure and rheology of complex
fluids}, Oxford University Press, New York.

\bibitem[Lauridsen, Twardos \& Dennin (2002)]{Lauridsen2002} \textsc{J. Lauridsen, M. Twardos \& M.
Dennin} 2002 Shear-induced stress relaxation in a two-dimensional
wet foam. \emph{Phys. Rev. Lett.} \textbf{89}, 098303.

\bibitem[Lauridsen, Chanan \& Dennin (2004)]{Lauridsen2004} \textsc{J. Lauridsen, G. Chanan \& M. Dennin} 2004 \emph{Phys. Rev. Lett.}
Velocity profiles in slowly sheared bubble rafts \textbf{93},
018303.

\bibitem[Marmottant \emph{et al.}]{Graner} \textsc{P. Marmottant, B. Dollet, C. Raufaste \& F. Graner}
Observation and prediction of local rearrangements: Plasticity in
a flowing foam, submitted.

\bibitem[Mason, Bibette \& Weitz (1995)]{Mason1995} \textsc{T. G. Mason, J. Bibette \& D. A. Weitz} 1995
Elasticity of compressed emulsions. \emph{Phys. Rev. Lett.}
\textbf{75}, 2051--2054.

\bibitem[Mason, Bibette \& Weitz (1996)]{Mason1996} \textsc{T. G. Mason, J. Bibette \& D. A.
Weitz} 1996 Yielding and flow of monodisperse emulsions.
 \emph{J. Coll. Int. Sci.} \textbf{179}, 439--448.

\bibitem[Mitsoulis (2004)]{Mitsoulis2004} \textsc{E. Mitsoulis} 2004
On creeping drag flow of a viscoplatic fluid past a circular
cylinder: Wall effects. \emph{Chem. Eng. Sci.} \textbf{59},
789--800.

\bibitem[Picard \emph{et al.} (2004)]{Picard2004} \textsc{G. Picard, A. Adjari, F. Lequeux \& L.
Bocquet} 2004 Elastic consequences of a single plastic event: A
step towards the microscopic modeling of the flow of yield stress
fluid. \emph{Eur. Phys. J. E} \textbf{15}, 371--381.

\bibitem[Pratt \& Dennin (2003)]{Pratt2003} \textsc{E. Pratt \& M. Dennin} 2003
Nonlinear stress and fluctuation dynamics of sheared disordered
wet foam \emph{Phys. Rev. E} \textbf{67}, 054102.

\bibitem[Princen (1983)]{Princen1983} \textsc{H. M. Princen} 1983
Rheology of foams and highly concentrated emulsions.
 I. Elastic properties and yield stress of a cylindrical model system. \emph{J. Coll. Int. Sci.} \textbf{91},
160--175.

\bibitem[Princen (1985)]{Princen1985} \textsc{H. M. Princen} 1985
Rheology of foams and highly concentrated emulsions. II.
Experimental study of the yield stress and wall effects for
concentrated oil-in-water emulsions. \emph{J. Coll. Int. Sci.}
\textbf{105}, 150--171.

\bibitem{Raufaste} \textsc{C. Raufaste,
B. Dollet, S. Cox, F. Graner \& Y. Jiang} Yield drag in a
two-dimensional foam around a circular obstacle: Effect of fluid
fraction, in preparation.


\bibitem[Roquet \& Saramito (2003)]{Roquet2003} \textsc{N. Roquet \& P. Saramito}
2003 An adaptive finite element mathod for Bingham fluid flows
around a cylinder \emph{Comput. Methods Appl. Mech. Eng.}
\textbf{192}, 3317--3341.

\bibitem[Saint-Jalmes \& Durian (1999)]{Saint-Jalmes1999} \textsc{A. Saint-Jalmes \& D. J. Durian} 1999
 Vanishing elasticity
for wet foams: Equivalence with emulsions and role of
polydispersity. \emph{J. Rheol.} \textbf{43}, 1411--1422.

\bibitem[Sollich \emph{et al.} (1997)]{Sollich1997} \textsc{P. Sollich, F. Lequeux, P. H\'{e}braud \& M. E.
Cates} 1997 Rheology of soft glassy materials. \emph{Phys. Rev.
Lett.} \textbf{78}, 2020--2023.

\bibitem[Stamenovi\'c \& Wilson (1984)]{Stamenovic1984} \textsc{D. Stamenovi\'c \& T. A.
Wilson} 1984 The shear modulus of liquid foam. \emph{J. Appl.
Mech.} \textbf{51}, 229--231.

\bibitem[Takeshi \& Sekimoto (2005)]{Takeshi2005} \textsc{O. Takeshi \& K. Sekimoto} 2005
Internal stress in a model elastoplastic fluid. \emph{Phys. Rev.
Lett.} \textbf{95}, 108301.

\bibitem[Vaz \& Cox (2005)]{Vaz2005} \textsc{M. F. Vaz \& S. J. Cox} 2005 Two-bubble instabilities in quasi-two-dimensional foams.
\emph{Phil. Mag. Lett.}
\textbf{85}, 415--425.

\bibitem[Wang, Krishan \& Dennin (2006)]{Wang} \textsc{Y. Wang, K. Krishan \& M. Dennin}
2006 Impact of boundaries on velocity profiles in bubble rafts.
 \emph{Phys. Rev. E} \textbf{73}, 031401.

\bibitem[Weaire \& Hutzler (1999)]{Weaire1999} \textsc{D. Weaire \& S. Hutzler} 1999 {\em {The Physics of
Foams}}, Oxford University Press, Oxford.

\bibitem[Weaire, Janiaud \& Hutzler]{Weaire} \textsc{D. Weaire, \'E. Janiaud \& S. Hutzler}
Two-dimensional foam rheology with viscous drag, submitted,
\texttt{arXiv:cond-mat/0602021}.

\end{thebibliography}
\end{document}